\documentclass[a4paper]{article}

\usepackage{url,hyperref,lineno,microtype,siunitx,subcaption,booktabs}
\usepackage[onehalfspacing]{setspace}
 
\usepackage{natbib,amsmath,booktabs,graphicx,authblk}
\usepackage[left=3cm, right=3cm, top=3cm, bottom=3cm]{geometry}

\usepackage[english]{babel}

\usepackage{todo}

\usepackage{listings}
\usepackage{color}
\usepackage[warn]{textcomp}

\definecolor{comments}{RGB}{0,0,113}
\definecolor{red}{RGB}{160,0,0}
\definecolor{green}{RGB}{0,150,0}

\newcommand{\CodeSymbol}[1]{\textcolor{green}{#1}}
\newcommand\Authors{Werner Van Geit\,$^{1,*}$, Michael Gevaert\,$^{1}$,
	Giuseppe Chindemi\,$^{1}$, Christian R{\"o}ssert\,$^{1}$,
	Jean-Denis Courcol\,$^{1}$, Eilif Muller\,$^{1}$, Felix
	Sch{\"u}rmann\,$^{1}$, Idan Segev\,$^{3,4}$ and Henry Markram\,$^{1,2,*}$}

\newcommand\Address{
$^{1}$Blue Brain Project, \'Ecole Polytechnique F\'ed\'erale de Lausanne (EPFL)
Biotech Campus, Geneva, Switzerland \\
$^{2}$Laboratory of Neural Microcircuitry, Brain Mind Institute, \'Ecole Polytechnique F\'ed\'erale de Lausanne, Lausanne, Switzerland \\
$^{3}$Department of Neurobiology, Alexander Silberman Institute of Life Sciences, The Hebrew University of Jerusalem, Jerusalem, Israel \\
$^{4}$The Edmond and Lily Safra Centre for Brain Sciences, The Hebrew University of Jerusalem, Jerusalem, Israel \\
\quad \\
$*$ werner.vangeit@epfl.ch, henry.markram@epfl.ch
}

\begin{document}
\onecolumn

\title{BluePyOpt: Leveraging open source software and cloud
	infrastructure to optimise model parameters in neuroscience}

\author{\Authors} 

\affil{\Address}

\maketitle

\begin{abstract}
  
At many scales in neuroscience, appropriate mathematical models take the form of complex dynamical systems.
Parametrising such models to conform to the multitude of available experimental constraints is a global nonlinear optimisation problem with a complex fitness landscape, requiring numerical techniques to find suitable approximate solutions.
Stochastic optimisation approaches, such as evolutionary algorithms, have been shown to be effective, but often the setting up of such optimisations and the choice of a specific search algorithm and its parameters is non-trivial, requiring domain-specific expertise.
Here we describe BluePyOpt, a Python package targeted at the broad neuroscience community to simplify this task.
BluePyOpt is an extensible framework for data-driven model parameter optimisation that wraps and standardises several existing open-source tools.
It simplifies the task of creating and sharing these optimisations, and the associated techniques and knowledge.
This is achieved by abstracting the optimisation and evaluation tasks into various reusable and flexible discrete elements according to established best-practices.
Further, BluePyOpt provides methods for setting up both small- and large-scale optimisations on a variety of platforms, ranging from laptops to Linux clusters and cloud-based compute infrastructures.
The versatility of the BluePyOpt framework is demonstrated by working through three representative neuroscience specific use cases.

\tiny

\end{abstract}

\section{Introduction}

Advances in experimental neuroscience are bringing an increasing volume and variety of data, and inspiring the development of larger and more detailed models \citep{izhikevich,Merolla,nmc,biospaun}.
While experimental constraints are usually available for the emergent behaviours of such models, it is unfortunately commonplace that many model parameters remain inaccessible to experimental techniques.
The problem of inferring or searching for model parameters that match model behaviours to experimental constraints constitutes an inverse problem \citep{inversep}, for which analytical solutions rarely exist for complex dynamical systems, i.e. most mathematical models in neuroscience.
Historically, such parameter searches were done by hand tuning, but the advent of increasingly powerful computing resources has brought automated search algorithms that can find suitable parameters \citep{bhallabower, vanierbower, achard, korngreen, druckmann, neurofitter, biolcyberreview, huys, marder, hay, emoo, Svensson2012, optimizer, giffit, patchclampbook}.
While many varieties of search algorithms have been described and explored in the literature \citep{vanierbower, biolcyberreview, Svensson2012}, stochastic optimisation approaches, such as simulated annealing and evolutionary algorithms, have been shown to be particularly effective strategies for such parameter searches \citep{vanierbower, druckmann, korngreen, Svensson2012}.
Nevertheless, picking the right type of stochastic algorithm and setting it up correctly remains a non-trivial task requiring domain-specific expertise, and could be model and constraint specific \citep{biolcyberreview}.

With the aim of bringing widely applicable and state-of-the-art automated parameter search algorithms and techniques to the broad neuroscience community, we describe here a Python-based open-source optimisation framework, BluePyOpt, which is available on Github (see \citep{bluepyopt}), and is designed taking into account model optimisation experience accumulated during the Blue Brain Project \citep{druckmann, hay, nmc, nmcportal} and the ramp-up phase of the Human Brain Project.
The general purpose high-level programming language Python was chosen for developing BluePyOpt, so as to contribute to, and also leverage from the growing scientific and neuroscientific software ecosystem \citep{py4sci, pythonneuroscience}, including state-of-the-art search algorithm implementations, modelling and data access tools.

At its core, BluePyOpt is a framework providing a conceptual scaffolding in the form of an object-oriented application programming interface or API for constructing optimization problems according to established best-practices, while leveraging existing search algorithms and modelling simulators transparently ``under the hood''.
For common optimisation tasks, the user configures the optimisation by writing a short Python script using the BluePyOpt API.
For more advanced use cases, the user is free to extend the API for their own needs, potentially contributing these extensions back to the core library.
The latter is important for BluePyOpt APIs to remain broadly applicable and state-of-the-art, as best-practices develop for specific problem domains, mirroring the evolution that has occured for neuron model optimization strategies \citep{bhallabower, hay}.

Depending on the complexity of the model to be optimised, BluePyOpt optimisations can require significant computing resources.
The systems available to neuroscientists in the community can be very heterogeneous, and it is often difficult for users to set up the required software.
BluePyOpt therefore also provides a novel cloud configuration mechanism to automate setting up the required environment on a local machine, cluster system, or cloud service such as Amazon Web Services.

To begin, this technical report provides an overview of the conceptual framework and open-source technologies used by BluePyOpt, followed by a presentation of the software architecture and API of BluePyOpt.
Next, three concrete use cases are elaborated in detail, showing how the BluePyOpt APIs, concepts and techniques can be put to use by potential users.
The first use case is an introductory example demonstrating the optimisation of a single compartmental neuron model with two Hodgkin-Huxley ion channels.
The second use case shows a BluePyOpt-based state-of-the-art optimisation of a morphologically detailed thick-tufted layer 5 pyramidal cell model of the type used in a recent \emph{in silico} reconstruction of a neocortical microcircuit \citep{nmc}.
The third use case demonstrates the broad applicability of BluePyOpt, showing how it can also be used to optimise parameters of synaptic plasticity models.

\section{Concepts}

The BluePyOpt framework provides a powerful tool to optimise models in the field of neuroscience, by combining several established Python-based open-source software technologies.
In particular, BluePyOpt leverages libraries providing optimisation algorithms, parallelisation, compute environment setup, and experimental data analysis.
For numerical evaluation of neuroscientific models, many open-source simulators with Python bindings are available for the user to chose from.
The common bridge allowing BluePyOpt to integrate these various softwares is the Python programming language, which has seen considerable uptake and a rapidly growing domain-specific software ecosystem in the neuroscience modelling community in recent years \citep{pythonneuroscience}.
Python is recognized as a programming language which is fun and easy to learn, yet also attractive to experts, meaning that novice and advanced programmers alike can easily use BluePyOpt, and contribute solutions to neuroscientific optimisation problems back to the community.

\begin{figure}
	\begin{center}
		\includegraphics[scale=.75,keepaspectratio]{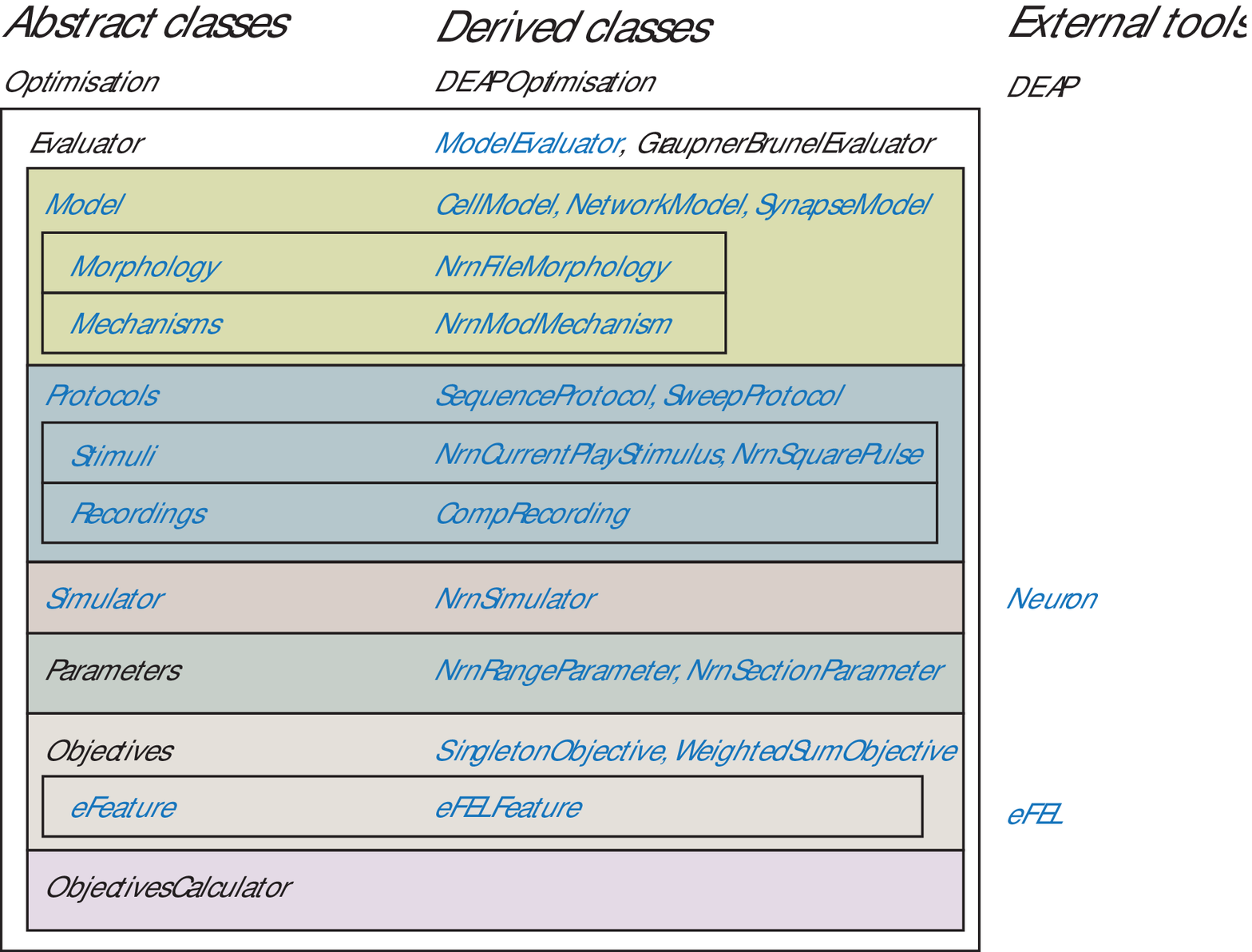}
	\end{center}
	\textbf{\refstepcounter{figure}\label{fig:classhierarchy} Figure
		\arabic{figure}.}{Hierarchy of the most important classes in BluePyOpt. 
		Ephys abstraction layer in \emph{blue}.}
\end{figure}

BluePyOpt was developed using an object oriented programming model.
Figure \ref{fig:classhierarchy} shows an overview of the class hierarchy of BluePyOpt.  
In its essence, the BluePyOpt object model defines the \emph{Optimisation} class which applies a search algorithm to an \emph{Evaluator} class.
Both are \textit{abstract classes}, meaning they define the object model, but not the implementation.
Taking advantage of Pythonic \textit{duck typing}, the user can then choose from a menu of implementations, \textit{derived classes}, or easily define their own implementations to meet their specific needs.  This design makes BluePyOpt highly versatile, while keeping the API complexity to a minimum.
The choice of algorithm and evaluator is up to the user, but many are already provided for various use cases.
For many common use cases, these are the only classes users are required to instantiate.

For neuron model optimizations in particular, BluePyOpt provides further classes to support feature-based multi-objective optimizations using NEURON, as shown in Figure \ref{fig:classhierarchy}.   
Classes \textit{CellModel}, \textit{Morphology}, \textit{Mechanisms}, \textit{Protocol}, \textit{Stimuli}, \textit{Recordings}, \textit{Location} are specific to setting up neuron models and assessing their input-output properties.
Other classes \textit{Objectives} and \textit{eFeature} are more generally applicable, with \textit{derived classes} for specific use cases, e.g. \textit{eFELFeature} provides features extracted from voltage traces using the open-source eFEL library discussed below.
They define features and objectives for feature-based multi-objective optimization, a best-in-class stochastic optimization strategy \citep{druckmann, druckmann_plos, hay}.
We generally recommend it as the first algorithm to try for a given problem domain.
For example, the third example for the optimisation of synaptic plasticity models also employs this strategy.

In the sub-sections to follow, an overview is provided for the various software components and the manner in which BluePyOpt integrates them.

\subsection{Optimisation algorithms}

Multiobjective evolutionary algorithms have been shown to perform well to optimise parameters of biophysically detailed multicompartmental neuron models \citep{druckmann, hay}.
To provide optimisation algorithms, BluePyOpt relies on a mature Python library, Distributed Evolutionary Algorithms in Python (DEAP), which implements a range of such algorithms \citep{DEAP_JMLR2012}.
The advantage of using this library is that it provides many useful features out of the box, and it is mature, actively maintained and well documented.
DEAP provides many popular algorithms, such as Non-dominated Sorting Genetic Algorithm-II \citep{nsga2}, Covariance Matrix Adaptation Evolution Strategy \citep{cma}, and Particle Swarm Optimisation \citep{pso}.
Moreover, due to its extensible design, implementing new search algorithms in DEAP is straight-forward.
Historically, the Blue Brain Project has used a \textit{C} implementation of the Indicator Based Evolutionary Algorithm \textit{IBEA} to optimise the parameters of biophysically detailed neuron models \citep{ZK2004a, pisa, nmc}, as this has been shown to have excellent convergence properties for these problems \citep{schmucker}.
Case in point, we implemented a version of IBEA for the DEAP framework, so this algorithm is consequently available to be used in BluePyOpt.

Moreover, DEAP is highly versatile, whereby most central members of its class hierarchy, such as individuals and operators, are fully customizable with user defined implementations.
Classes are provided to keep track of the \textit{Pareto Front} or the \textit{Hall-of-Fame} of individuals during evolution.
Population statistics can be recorded in a logbook, and the genealogy between individuals can be saved, analysed and visualised.
In addition, \textit{checkpointing} can be implemented in DEAP by storing the algorithm's state in a Python pickle file for any generation, as described in DEAP's documentation \citep{deapdocs}.

Although the use cases below use DEAP as a library to implement the search algorithm, it is worth noting that BluePyOpt abstracts the concept of a search algorithm.
As such, it is entirely possible to implement algorithms that are independent of DEAP, or that use other third-party libraries.

\subsection{Simulators}

To define a BluePyOpt optimisation, the user must provide an evaluation function which maps model parameters to a fitness score.
It can be a single Python function that maps the parameters to objectives by solving a set of equations, or a function that uses an external simulator to evaluate a complex model under multiple scenarios.
For the latter, the only requirement BluePyOpt imposes is that it can interact with the external simulator from within Python.
Often, this interaction is implemented through Python modules provided by the user's neuroscientific simulator of choice, as is the case for many simulators in common use, including NEURON \citep{neuronpython}, NEST \citep{pynest,cynest}, PyNN \citep{pynn}, BRIAN \citep{briansim}, STEPS \citep{pysteps}, and MOOSE \citep{pymoose}.
Otherwise, communication through shell commands and input/output files is also possible, so long as an interface can be provided as a Python class.

\subsection{Feature Extraction}

For an evaluation function to compute a fitness score from simulator output, the resulting traces must be compared against experimental constraints.
Voltage recordings obtained from patch clamp experiments are an example of experimental data that can be used as a constraint for neuron models.
From such recordings the neuroscientist can deduce many interesting values, like the input resistance of the neuron, the action potential characteristics, firing frequency etc.
To standardise the way these values are measured, the Blue Brain Project has released the Electrophysiology Feature Extract Library (eFEL) \citep{efel}, also as open-source software.
The core of this library is written in C++, and a Python wrapper is provided.
BluePyOpt can interact with eFEL to compute a variety of features of the voltage response of neuron models.
A fitness score can then be computed by some distance metric comparing the resulting model features to their experimental counterparts.
As we will see for the last example in this article, a similar approach can also be taken for other optimisation problem domains.

\subsection{Parallelisation}

Optimisations of the parameters of an evaluation function typically require the execution of this function repeatedly.
For a given optimisation integration step, such executions are often in the hundreds (scaling e.g. with evolutionary algorithm population size), are compute bound, and are essentially independent, making them ripe for parallelisation.
Parallelisation of the optimisation can be performed in several ways.
DEAP provides an easy way to evaluate individuals in a population on several cores in parallel.
The user need merely provide an implementation of a \emph{map} function.
In its simplest form, this function can be the Python serial \emph{map} in the standard library, or the parallel \emph{map} function in the multiprocessing module to leverage local hardware threads.
To parallelise over a large cluster machine, the DEAP developers encourage the use of the SCOOP \citep{scoop} map function.
SCOOP is a library that builds on top of ZeroMQ \citep{zeromq}, which provides a socket communication layer to distribute the computation over several computers.
Other map functions and technologies can be used like MPI4Py \citep{mpi4py} or iPython ipyparallel package \citep{ipython}.
Moreover, parallelisation doesn't necessarily have to happen at the population level.
Inside the evaluation of individuals, map functions can also be used to parallelise over stimulus protocols, feature types, etc., however for the problem examples presented here, such an approach wouldn't make good use of anything more than 10 to 20 cores.

\subsection{Cloud}

To increase the throughput of optimisations, multiple computers can be used to parallelise the work.
Such a group of computers can be composed of machines in a cluster, or they can be obtained from a cloud provider like Amazon Web Services, Rackspace Public Cloud, Microsoft Azure, Google Compute Engine, or the Neuroscience Gateway portal \citep{nsgportal}.

These and other cloud providers allow for precise allocation of numbers of machines and their storage, compute power and memory.
Depending on the needs and resources of an individual or organization, trade-offs can be made on how much to spend versus how fast the results are needed.

Setting up a cluster or cloud environment with the correct software requirements is often complicated and error prone: Each environment has to be exactly the same, and scripts and data need to be available in the same locations.
To ease the burden of this configuration, BluePyOpt includes Ansible \citep{ansible} configuration scripts for setting up a local test environment (on one machine, using Vagrant \citep{vagrant}), for setting up a cluster with a shared file system, or for provisioning and setting up an Amazon Web Services cluster.

Ansible is open-source software that allows for reproducible environments to be created and configured from simple textual descriptions called 'Playbooks'.
These Playbooks encapsulate the discrete steps needed to create an environment, and offer extra tools to simplify things like package management, user creation and key distribution.
Furthermore, when a Playbook is changed and run against an already existing environment, only the changes necessary will be applied.
Finally, Ansible has the advantage over other systems, like Puppet \citep{puppet} and Chef \citep{chef}, that nothing except a Python interpreter needs to be installed on the target machine and all environment discovery and configuration is performed through SSH from the machine on which Ansible is run.
This decentralized system means that a user can use Ansible to setup an environment in their home directory on a cluster, without intervention from the system administrators.

\section{Software Architecture}

\begin{figure}
\begin{center}
\includegraphics[scale=.5,keepaspectratio]{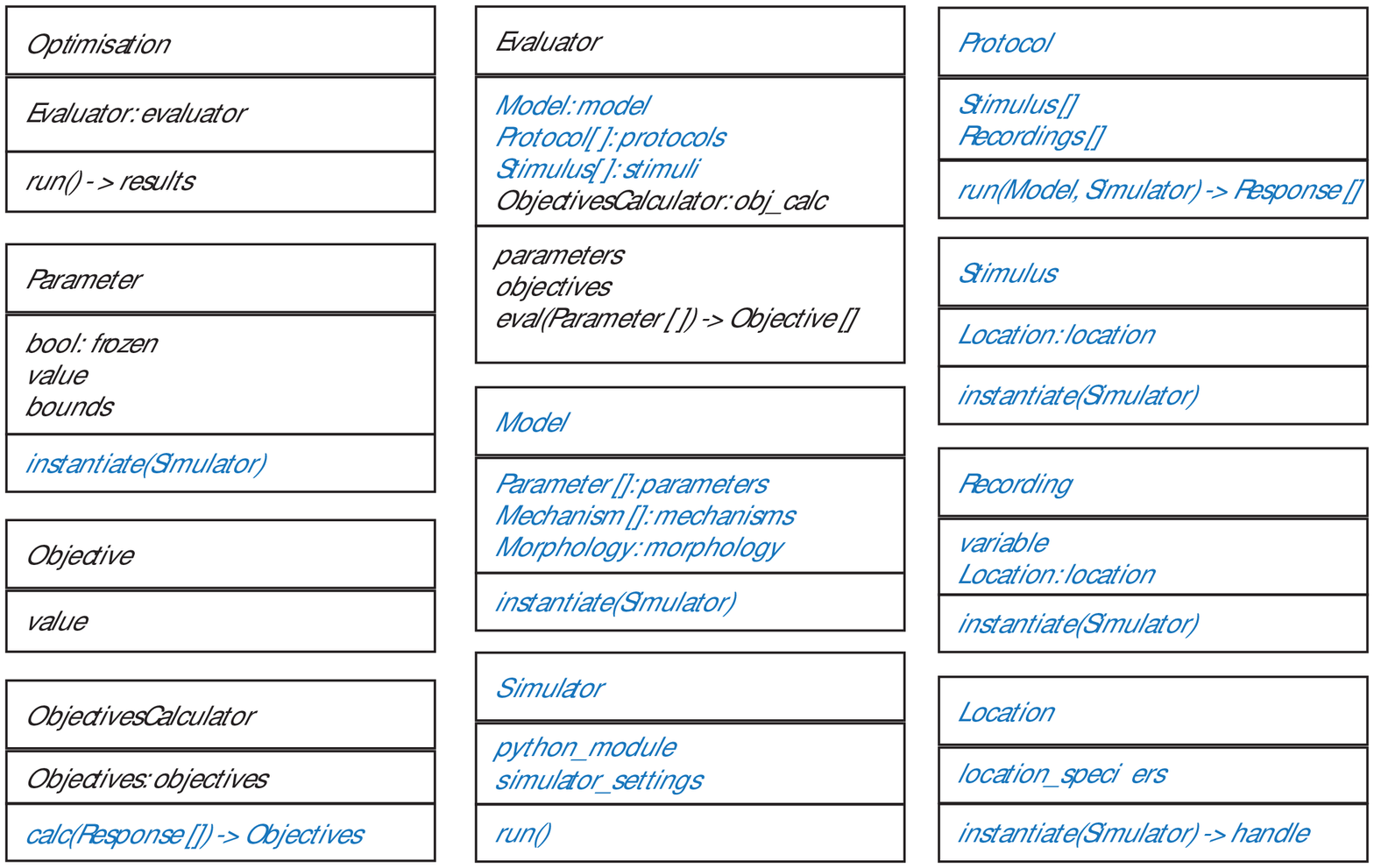}
\end{center}
 \caption{General structure of most important classes. Every box represents a class. In every box the top panel is the name, the middle panel the most important fields and the bottom panel the most important methods. 
Ephys abstraction layer in \emph{blue}.}\label{fig:classes}
\end{figure}

\begin{figure}
\begin{center}
\includegraphics[scale=.75,keepaspectratio]{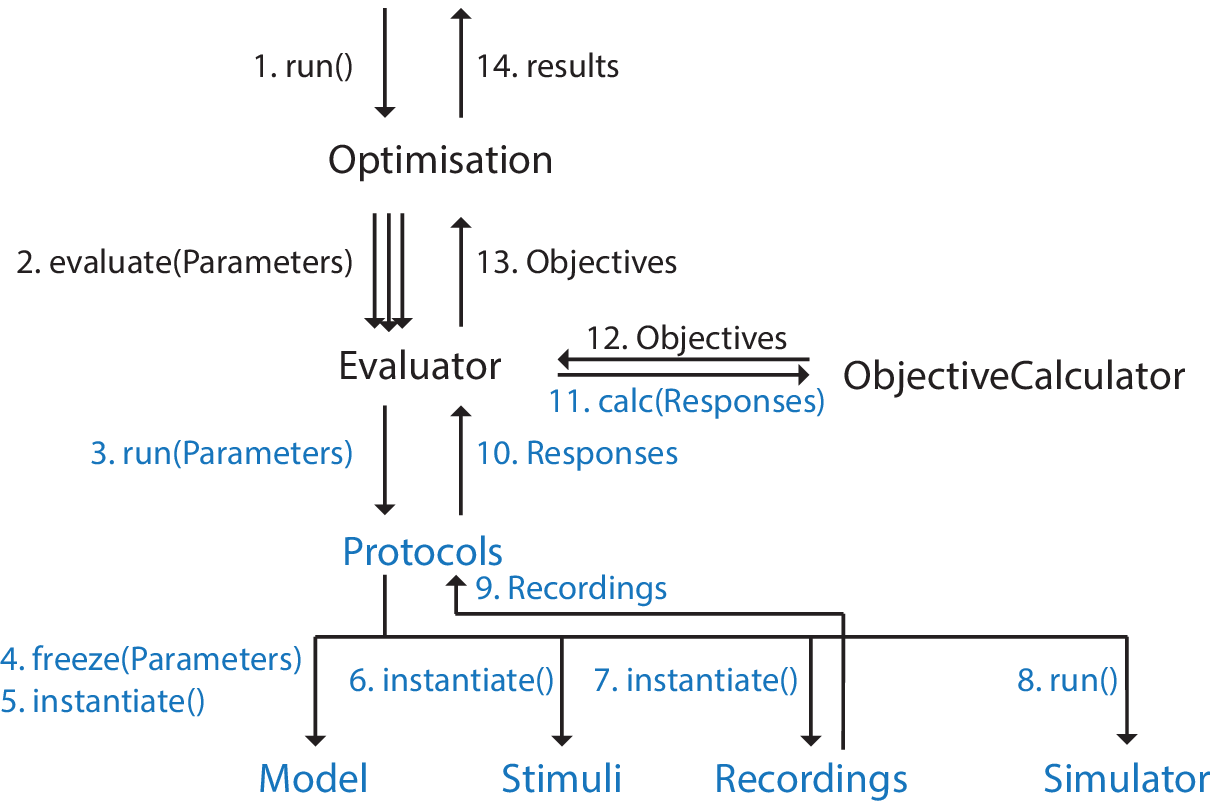}
\end{center}
\caption{Graph representing control flow in BluePyOpt. Ordering is clarified by the numbers. Arrow labels that contain parentheses represent function calls, the other labels data being returned. This figure is meant to give a high level description of the control flow, not all function calls and intermediate objects are included.
Ephys abstraction layer in \emph{blue}.}\label{fig:workflow} 
\end{figure}

The BluePyOpt software architecture follows an object oriented programming model, whereby the various concepts of the software are modularised into cleanly seperated and well defined classes which interact, as defined in a class hierarchy (Figure \ref{fig:classhierarchy}) and object model (Figure \ref{fig:classes}) show to define the program control flow, as shown in Figure \ref{fig:workflow}.
In what follows, the role of each class and how it relates to and interacts with other classes in the hierarchy is described.

\subsection{Optimisation abstraction layer}

At the highest level of abstraction, the BluePyOpt \textit{API} contains the classes \emph{Optimisation} and \emph{Evaluator} (Figure \ref{fig:classes}). 
An \emph{Evaluator} object defines an evaluation function that maps \emph{Parameters} to \emph{Objectives}.
The \emph{Optimisation} object accepts the \emph{Evaluator} as input, and runs a search algorithm on it to find the parameter values that generate the best objectives.

The task of the search algorithm is to find the parameter values that correspond to the best objective values. 
Defining 'best' is left to the specific implementation.
As in the use cases below, this could be a weighted sum of the objectives or a multiobjective front in a multidimensional space.

The \emph{Optimisation} class allows the user to control the settings of the search algorithm. 
In case of IBEA, this could be the number of individuals in the population, the mutation probabilities, etc.

\subsection{EPhys model abstraction layer}
On a different level of abstraction, we have classes that are tailored for electrophysiology (ephys) experiments and can be used inside the \emph{Evaluator}. The ephys model layer provides an abstraction to the simulator, so that the person performing the optimisation doesn't have to have knowledge of the intricate details of the simulator.

A \emph{Protocol} is applied to a \emph{Model} for a certain set of \emph{Parameters}, generating a \emph{Response}.
An \emph{ObjectivesCalculator} is then used to calculate the \emph{Objectives} based on the \emph{Response} of the \emph{Model}.
All these classes are part of the \verb^bluepyopt.ephys^ package.

\subsubsection{Model}
By making a \emph{Model} an abstract class, we give users the ability to use our software for a broad range of use cases.
A \emph{Protocol} can attach \emph{Stimuli} and \emph{Recordings} to a \emph{Model}. When the \emph{Simulator} is then run, a \emph{Response} is generated for each of the \emph{Stimuli} for a given set of \emph{Model} parameter values.

Examples of broad subclasses are a \emph{NetworkModel}, \emph{CellModel} and \emph{SynapseModel}.
Specific subclasses can be made for different simulators, or assuming some level of similarity, the same model object can know how to instantiate itself in different simulators.
In the future, functionality could be added to import/export the model configuration from/to standard description languages like NeuroML \citep{neuroml} or NineML \citep{nineml}.

Particular parameters of a \emph{Model} can be in a frozen state. This means that their value is fixed, and won't be considered for optimisation.
This concept can be useful in multi-stage optimisation in which subgroups of a model are optimised in a sequential fashion.

Another advantage of this abstraction is that a \emph{Model} is a standalone entity that can be run outside of the \emph{Optimisation} and have exactly the same \emph{Protocols} applied to it, generating exactly the same \emph{Response}.
One can also apply extra \emph{Protocols} to assess how well the model generalises, or to perfom a sensitivity analysis.

\subsubsection{Simulator}
Every model simulator should have a subclass of \emph{Simulator}.
Objects of this type will be passed on to objects that are simulator aware, like the \emph{Model} and \emph{Stimuli} when their \emph{instantiate} method is invoked. 
This architecture allows e.g. the same model object to be run in different simulators.
Examples of functionality this class can provide are links to the Python module related to the simulator, the enabling of variable time step integration, etc. \emph{Simulators} also have \emph{run()} method that starts the simulation.

\subsubsection{Protocol}
A \emph{Protocol} is an object that elicits a \emph{Response} from a model.
In its simplest form it represents, for example, a step current clamp stimulus, but more complicated versions are possible, such as stimulating a set of cells in a network with an elaborate protocol and recording the response.
A \emph{Protocol} can also contain sub-protocols, providing a powerful mechanism to reuse components.

\subsubsection{Stimulus, Recording and Response}
The \emph{Stimulus} and \emph{Recording} objects, which are part of a \emph{Protocol} are applied to a model and are aware of the simulator used.
Subclasses of \emph{Stimulus} are concepts like current/voltage clamp, synaptic activation, etc.
Both of these classes accept a \emph{Location} specifier. Several \emph{Recording} objects can be combined in a \emph{Response} which can be analysed by an \emph{ObjectiveCalculator}.

\subsubsection{Location}
Specifying the location on a neuron morphology of a recording, stimulus or parameter in a simulator can be complicated.
Therefore we created an abstract class \emph{Location}.
As arguments the constructor accepts the location specification, e.g. in NEURON this could be a \emph{sectionlist} name and an index of the section, or it could point to a section at a certain distance from the soma.
Upon request, the object will return a reference to the object at the specified location, this could e.g be a NEURON section or compartment.
At a location, a variable can be set or recorded by a \emph{Parameter} or \emph{Recording}, respectively.

\subsubsection{ObjectivesCalculator, eFeature}

The \emph{ObjectivesCalculator} takes the \emph{Response} of a \emph{Model} and calculates the objective values from it.
When using ephys recordings, one can use the eFEL library to extract \emph{eFeatures}.
Examples of these eFeatures are spike amplitudes, steady state voltages, etc.
The values of these eFeatures can then be compared with experimental data values, and a score can be calculated based on the difference between model and experiment.

\section{Example Use Cases}

To provide hands on experience how real-world optimisations can be developed using the BluePyOpt API, this section provides step-by-step guides for three use-cases.
The first is a single compartmental neuron model optimisation, the second is an optimisation of a state-of-the-art morphologically detailed neuron model, and the third is an optimisation of a synaptic plasticity model.
All examples to follow assume NEURON default units, i.e. \SI{}{\milli\second}, \SI{}{\milli\volt}, \SI{}{\nano\ampere}, \SI{}{\micro\farad\per\cm\squared}, etc. \citep{neuronunits}.

\subsection{Single compartmental model}

The first use case shows how to set up an optimisation of single compartmental neuron model with two free parameters: The maximal conductances of the sodium and potassium Hodgkin-Huxley ion channels.
This example serves as an introduction for the user to the programming concepts in BluePyOpt.
It uses the NEURON simulator as backend.

First we need to import the top-level \verb^bluepyopt^ module.
This example will also use BluePyOpt's electrophysiology features, so we also need to import the \verb^bluepyopt.ephys^ subpackage.

\begin{lstlisting}
import bluepyopt as bpop
import bluepyopt.ephys as ephys
\end{lstlisting}

Next we load a morphology from a file.
By default a morphology in NEURON has the following \emph{sectionlists}: somatic, axonal, apical and basal. We create a \emph{Location} object (specifically, a \emph{NrnSecListLocation} object) that points to the somatic sectionlist.
This object will be used later to specify where mechanisms are to be added etc.

\begin{lstlisting}
morph = ephys.morphologies.NrnFileMorphology('simple.swc')
somatic_loc = ephys.locations.NrnSeclistLocation('somatic', seclist_name='somatic')
\end{lstlisting}

Now we can add ion channels to this morphology.
First we add the default NEURON Hodgkin-Huxley mechanism to the soma, as follows.

\begin{lstlisting}
hh_mech = ephys.mechanisms.NrnMODMechanism(
	name='hh',
	prefix='hh',
	locations=[somatic_loc])
\end{lstlisting}

The \emph{name} argument can be chosen by the user, and should be unique across mechanisms.
The \emph{prefix} argument string should correspond to the SUFFIX field in the NEURON NMODL description file \citep{neuronbook} of the channel.  The \emph{locations} argument specifies which sections the mechanism are to be added to.

Next we need to specify the parameters of the model.
A parameter can be in two states: frozen and not-frozen.
When a parameter is frozen it has an exact known value, otherwise it has well-defined bounds but the exact value is not known yet.
The parameter for the capacitance of the soma will be a frozen value.

\begin{lstlisting}
cm_param = ephys.parameters.NrnSectionParameter(
	name='cm',
	param_name='cm',
	value=1.0,
	locations=[somatic_loc],
	frozen=True)
\end{lstlisting}

The two parameters that represent the maximal conductance of the sodium and potassium channels are to be optimised, and are therefore specified as \verb^frozen=False^, i.e. not-frozen, and bounds for each are provided with the \emph{bounds} argument.

\begin{lstlisting}
gnabar_param = ephys.parameters.NrnSectionParameter(
	name='gnabar_hh',
	param_name='gnabar_hh',
	locations=[somatic_loc],
	bounds=[0.05, 0.125],
	frozen=False)
gkbar_param = ephys.parameters.NrnSectionParameter(
	name='gkbar_hh',
	param_name='gkbar_hh',
	bounds=[0.01, 0.075],
	locations=[somatic_loc],
	frozen=False)
\end{lstlisting}

To create the cell template, we pass all these objects to the constructor of the model.

\begin{lstlisting}
simple_cell = ephys.cellmodels.NrnCellModel(
	name='simple_cell',
	morph=morph,
	mechs=[hh_mech],
	params=[cm_param, gnabar_param, gkbar_param])
\end{lstlisting}

To optimise the parameters of the cell, we further need to create a \emph{CellEvaluator} object.
This object needs to know which protocols to inject, which parameters to optimise, and how to compute a score, so we'll first create objects that define these aspects.

A protocol consists of a set of stimuli and a set of responses (i.e. recordings).
These responses will later be used to calculate the score of the specific model parameter values.
In this example, we will specify two stimuli, two square current pulses delivered at the soma with different amplitudes.
To this end, we first need to create a location object for the soma.

\begin{lstlisting}
soma_loc = ephys.locations.NrnSeclistLocation(
	name='soma',
	seclist_name='somatic',
	sec_index=0,
	comp_x=0.5)
\end{lstlisting}

For each step in the protocol, we add a stimulus (\emph{NrnSquarePulse}) and a recording (\emph{CompRecording}) in the soma.

\begin{lstlisting}
sweep_protocols = {}
for protocol_name, amplitude in [('step1', 0.01), ('step2', 0.05)]:
	stim = ephys.stimuli.NrnSquarePulse(
		step_amplitude=amplitude, 
		step_delay=100, 
		step_duration=50, 
		location=soma_loc,
		total_duration=200)
	rec = ephys.recordings.CompRecording(
	name='%s.soma.v' % protocol_name,
	location=soma_loc,
	variable='v')
	protocol = ephys.protocols.SweepProtocol(protocol_name, [stim], [rec])
	sweep_protocols[protocol.name] = protocol
\end{lstlisting}

The \verb^step_amplitude^ argument of the NrnSquarePulse specifies the amplitude of the current pulse, and \verb^step_delay^, \verb^step_duration^, and \verb^total_duration^ specify the start time, length and total simulation time.
Finally, we create a combined protocol that encapsulates both current pulse protocols.

\begin{lstlisting}
twostep_protocol = ephys.protocols.SequenceProtocol('twostep', protocols=sweep_protocols)
\end{lstlisting}

Now to compute the model score that will be used by the optimisation algorithm, we define objective objects.
For this example, our objective is to match the eFEL ``Spikecount'' feature to specified values for both current injection amplitudes.
In this case, we will create one objective per feature.

\begin{lstlisting}
efel_feature_means = {'step1': {'Spikecount': 1}, 'step2': {'Spikecount': 5}}

objectives = []

for protocol_name, protocol in protocols.iteritems():
	stim_start = protocol.stimuli[0].step_delay
	stim_end = stim_start + protocol.stimuli[0].step_duration
	for efel_feature_name, mean in
	efel_feature_means[protocol_name].iteritems():
		feature_name = '%s.%s' % (protocol_name, efel_feature_name)
		feature = ephys.efeatures.eFELFeature(
			feature_name,
			efel_feature_name=efel_feature_name,
			recording_names={'': '%s.soma.v' % protocol_name},
			stim_start=stim_start,
			stim_end=stim_end,
			exp_mean=mean,
			exp_std=0.05 * mean)
		objective = ephys.objectives.SingletonObjective(
			feature_name,
			feature)
		objectives.append(objective)
\end{lstlisting}

We then pass these objective definitions to a \emph{ObjectivesCalculator} object, calculate the total scores from a protocol response.

\begin{lstlisting}
obj_calc = ephys.scorecalculators.ObjectivesCalculator(objectives)
\end{lstlisting}

Finally, we can combine everything together into a \emph{CellEvaluator}.
The CellEvaluator constructor has a field \emph{param\_names} which contains the (ordered) list of names of the parameters that are used as input (and will be fitted later on).

\begin{lstlisting}
cell_evaluator = ephys.evaluators.CellEvaluator(
	cell_model=simple_cell,
	param_names=['gnabar_hh', 'gkbar_hh'],
	fitness_protocols=protocols,
	fitness_calculator=obj_calc)
\end{lstlisting}

Now that we have a cell template and an evaluator for this cell, the \emph{Optimisation} object can be created and run.

\begin{lstlisting}
optimisation = bpop.optimisations.DEAPOptimisation(
	evaluator=cell_evaluator,
	offspring_size = 100)

final_pop, hall_of_fame, logs, hist = optimisation.run(max_ngen=10)
\end{lstlisting}

\begin{figure}
	\begin{minipage}[b]{1.\linewidth}
		\begin{center}
			\includegraphics[scale=.4,keepaspectratio]{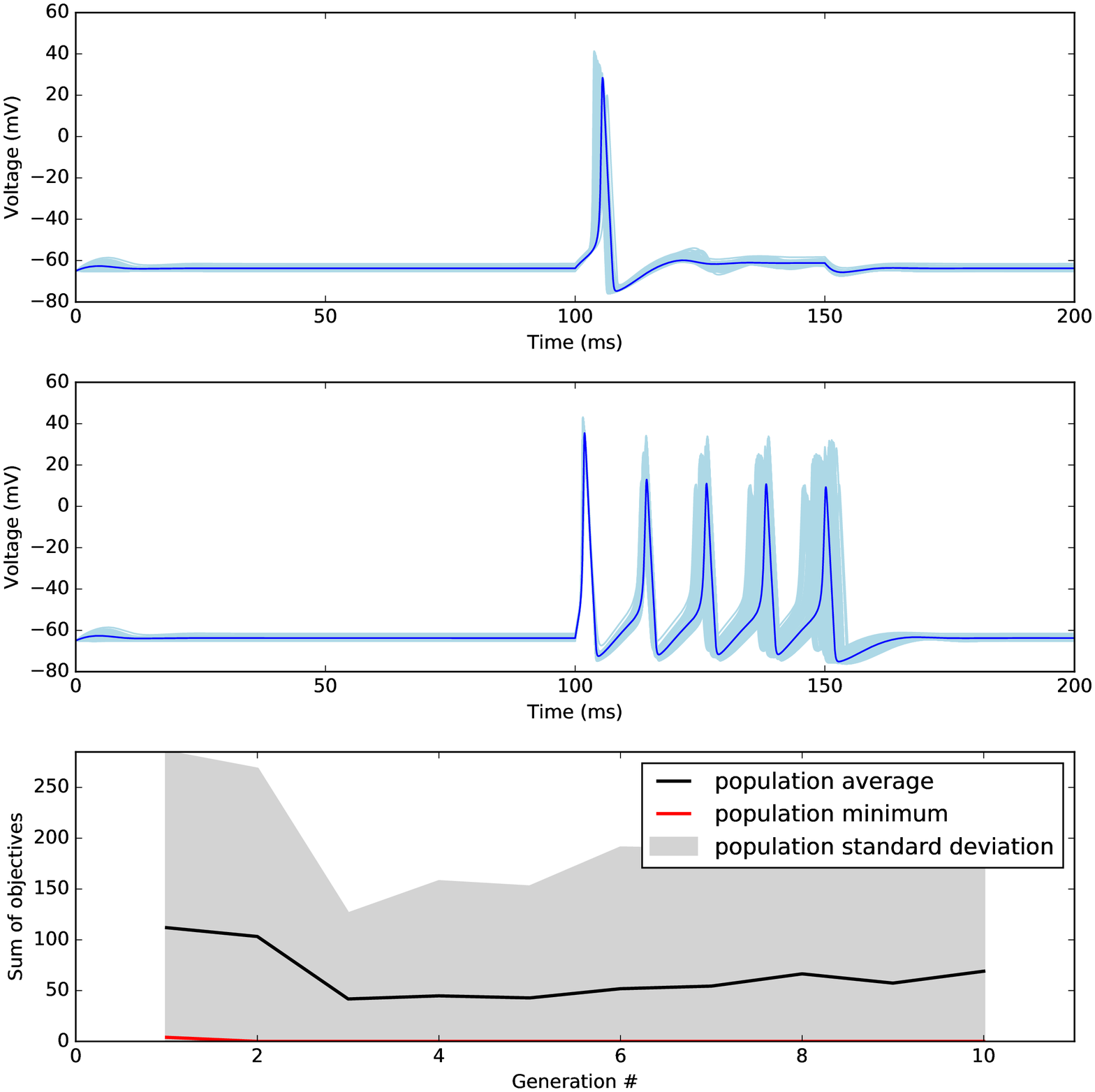}
		\end{center}
		\subcaption{\emph{Top plots}: In \emph{light blue}, voltage traces recording during the two different step current injection for all the individuals found that have objectives sum equal to zero. In \emph{dark blue}, an example of ones of these individuals. 
	The target objectives of \emph{Step1} and \emph{Step2} were 1 and 5 action potentials respectively. 
	\emph{Bottom plot}: Evolution of minimal objectives sum during the 10 generations of the evolutionary algorithm.}
	\end{minipage}
	\medskip	
	\begin{minipage}[b]{1.\linewidth}
		\begin{center}
			\includegraphics[scale=.5,keepaspectratio]{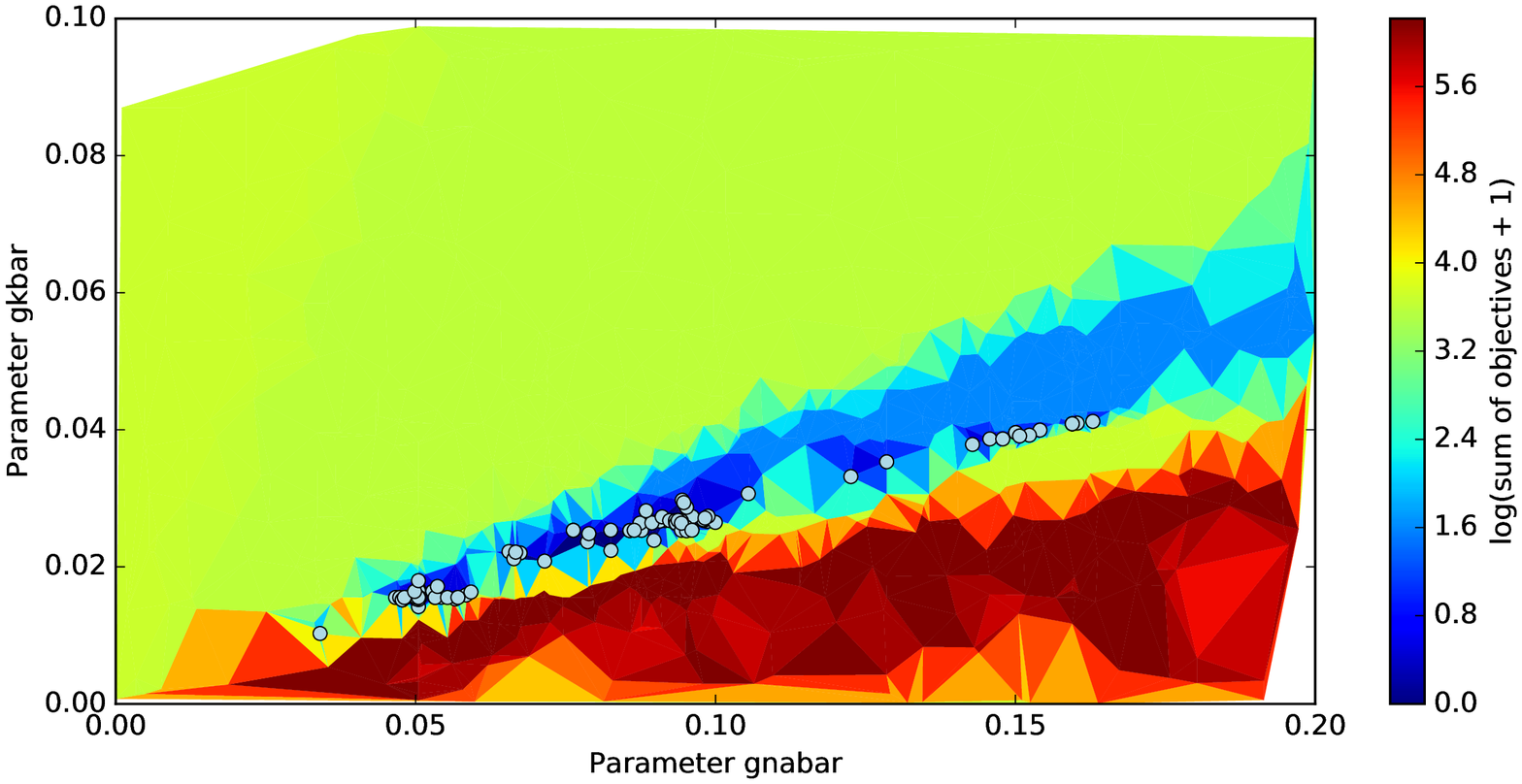}
		\end{center}
		\subcaption{Triangular grid plot of the parameter space. 
 		Every point of the grid is a point where the algorithm evaluated an individual. 
 		X- and Y-axis represent the values of the sodium and potassium maximal conductance respectively (units \SI{}{\siemens\per\cm\squared}).
		The color represents the average natural logarithm of the objectives sum of every triangle's three points. 
		Circles represent the solutions with an objectives sum of 0.}\label{fig:simplecell_trip}
	\end{minipage}

	\caption{Results of the single compartmental model optimisation.}\label{fig:simplecell}
\end{figure}

After a short time, the optimisation returns us the final population, the hall of fame, a logbook and an object containing the history of the population during the execution of the algorithm.
Figure \ref{fig:simplecell} shows the results in a graphical form.

\subsection{Neocortical pyramidal cell}

\begin{figure}
	\begin{minipage}[b]{.45\linewidth}
		\begin{center}
 		\includegraphics[scale=.35,keepaspectratio]{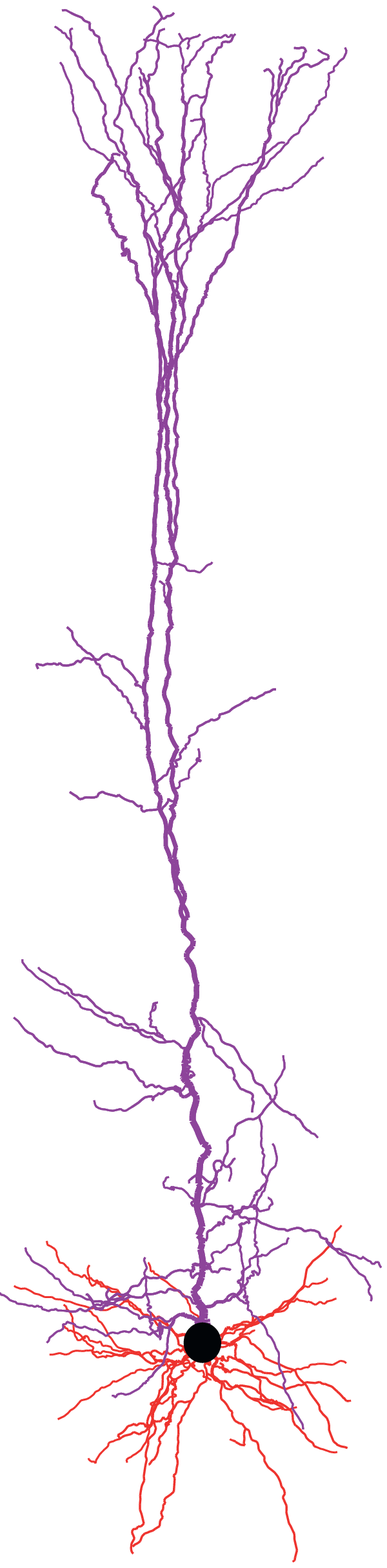}
 		\end{center}
 		\subcaption{Morphological reconstruction of L5PC used in the model obtained from the NMC portal \citep{nmcportal}.}\label{fig:l5pc_release_morph}
 	\end{minipage}
 	\hspace{0.05\linewidth}
	\begin{minipage}[b]{.45\linewidth}
		\begin{center}
		\includegraphics[scale=.35,keepaspectratio]{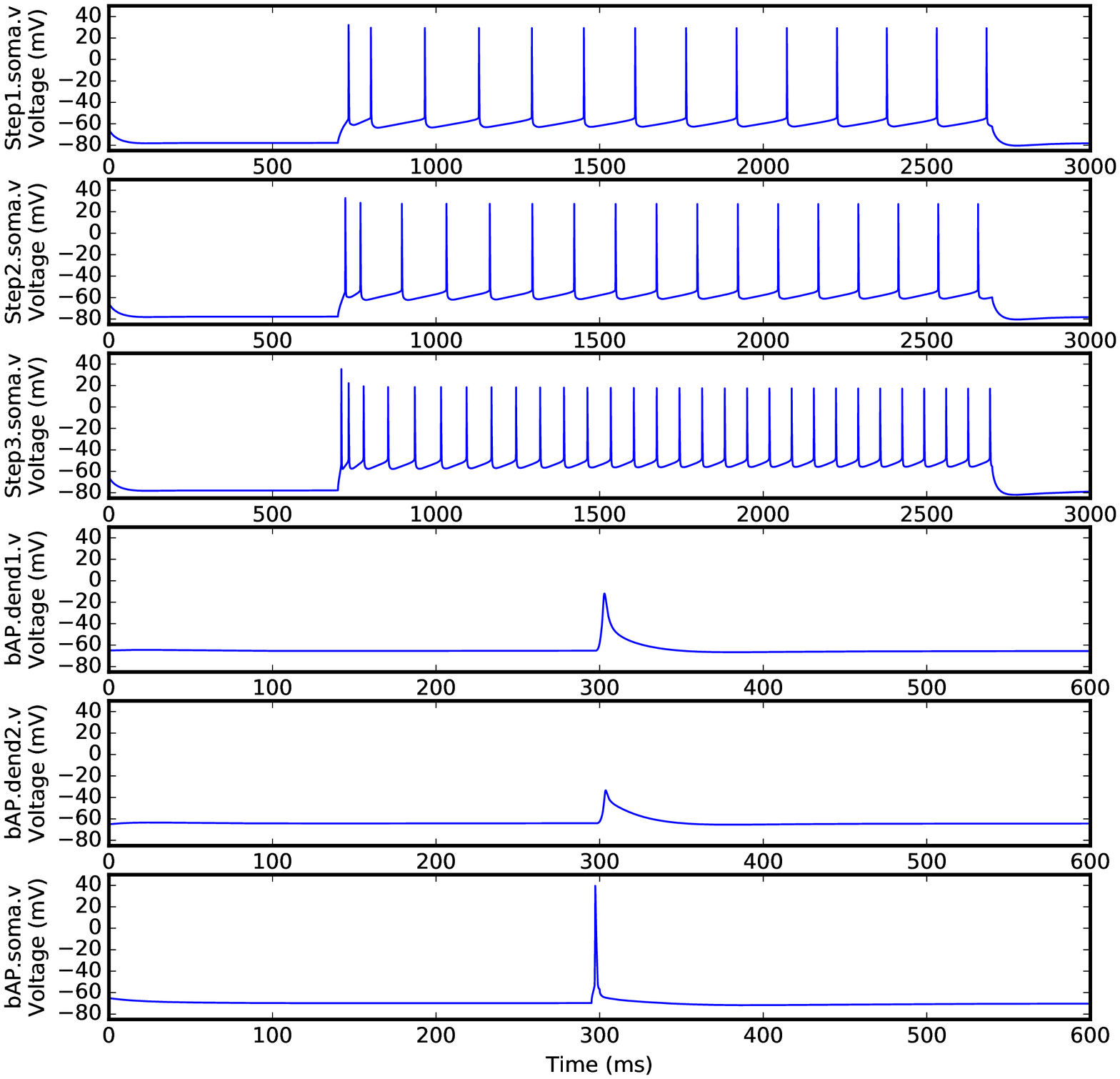}
		\end{center}
 		\subcaption{Voltage traces recorded in soma and dendrites (dend1 \SI{660}{\micro\metre}, dend2 \SI{800}{\micro\metre} from soma in apical trunk). }\label{fig:l5pc_release_model_responses}
 	\end{minipage}
 	
 	\medskip	
 	
	\begin{minipage}[b]{.45\linewidth}
		\begin{center}
		\includegraphics[scale=.35,keepaspectratio]{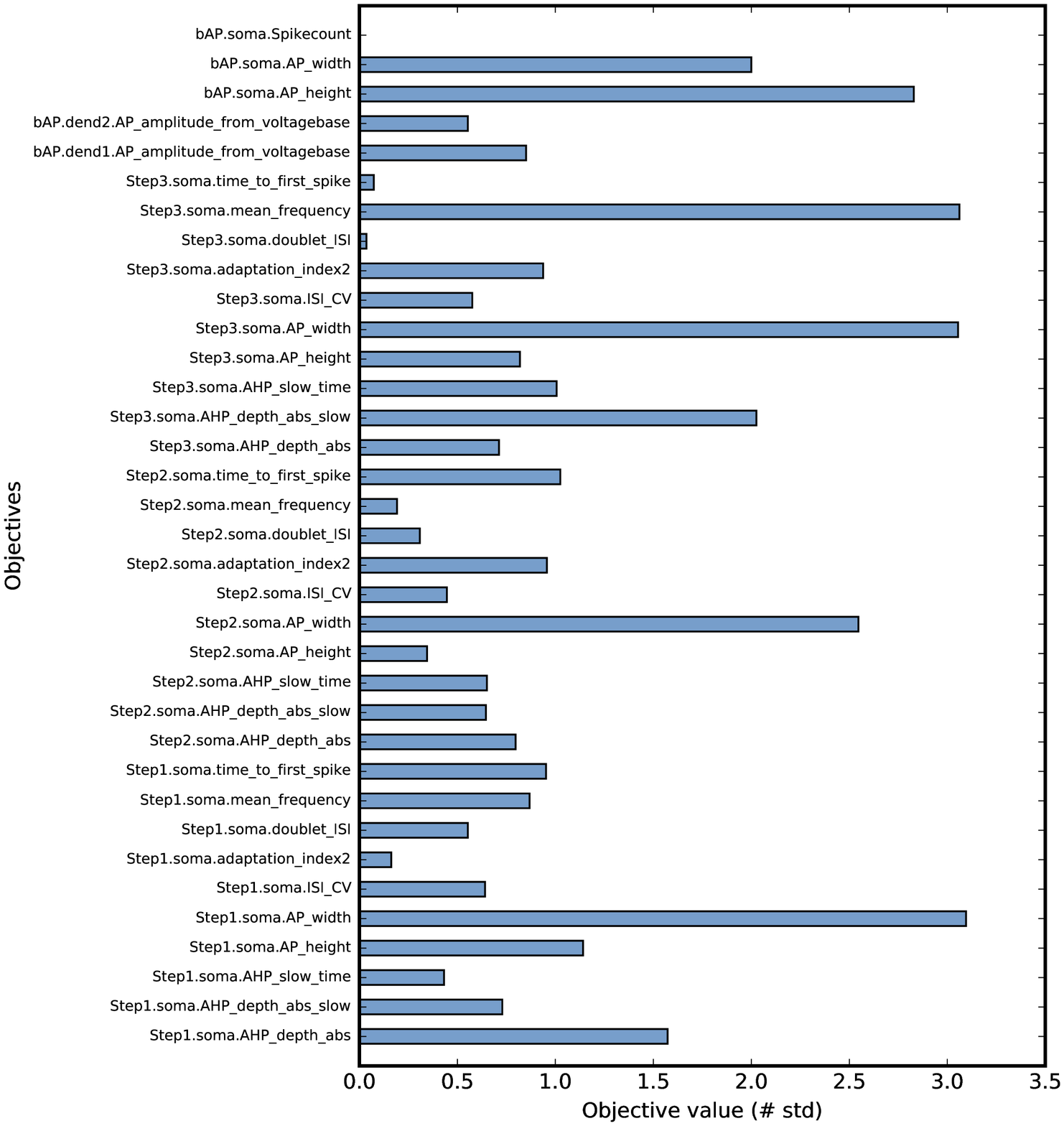}
		\end{center}
 		\subcaption{Objective scores for the model calculated based on experimental mean and standard deviation.}\label{fig:l5pc_release_model}
 	\end{minipage}

 	\caption{L5PC model as simulated by BluePyOpt with parameter values from \citep{nmc}}\label{fig:l5pc_release}
\end{figure}

Our second use case is a more complex example demonstrating the optimisation of a morphologically detailed model of a thick-tufted layer 5 pyramidal cell (L5PC) from the neocortex (Fig. \ref{fig:l5pc_release_morph}).
This example uses a BluePyOpt port of the state-of-the-art methods for the optimisation of the L5PC model described in \cite{nmc}.
The original model is available online from the Neocortical Microcircuit Collaboration Portal \citep{nmcportal}.
Due to its complexity, we will not describe the complete optimisation script here.
The full code is available from the BluePyOpt website.
What we will do here is highlight the particularities of this model compared to the introductory single compartmental model optimisation.  As a first validation and point of reference, we ran the BluePyOpt model with its original parameter values from \citep{nmcportal}, as shown in Figure \ref{fig:l5pc_release_model_responses}.

For clarity, the code for setting the parameters, objective and optimisation algorithm is partitioned into separate modules.
Configuration values are stored and read from JavaScript Object Notation \textit{JSON} files.

\subsubsection{Parameters}
Evidently, the parameters of this model, as shown in Table \ref{paramtable}, far exceed in number those of the single compartmental use case.
The parameters marked as frozen are kept a constant throughout the optimisation.
The parameters to be optimised are the maximal conductances of the ion channels and two values related to the calcium dynamics.
The location of the parameters is based on sectionlist names, whereby sections are automatically assigned to the somatic, axonal, apical and basal sectionlists by NEURON when it loads a morphology.

\begin{table}
\textbf{\refstepcounter{table}\label{paramtable} Table \arabic{table}.}{ List of parameters for L5PC example. 
Optimised parameters with bounds are in upper part of the table, lower part lists the frozen parameters with their value. The parameter with \emph{exp} distribution is scaled for every morphological segment with the equation $-0.8696 + 2.087\cdot e^{0.0031\cdot d}$ with \emph{d} the distance of the segment to the soma.}

\begin{tabular}{lllllll}
	\toprule
		Location & Mechanism & Parameter name & Distribution & Units & Lower bound & Upper bound \\
		\midrule
		apical & NaTs2{\_}t & gNaTs2{\_}tbar & uniform & \SI{}{\siemens\per\cm\squared} & 0 & 0.04 \\
		apical & SKv3{\_}1 & gSKv3{\_}1bar & uniform & \SI{}{\siemens\per\cm\squared} & 0 & 0.04 \\
		apical & Im & gImbar & uniform & \SI{}{\siemens\per\cm\squared} & 0 & 0.001 \\
		axonal & NaTa{\_}t & gNaTa{\_}tbar & uniform & \SI{}{\siemens\per\cm\squared} & 0 & 4 \\
		axonal & Nap{\_}Et2 & gNap{\_}Et2bar & uniform & \SI{}{\siemens\per\cm\squared} & 0 & 4 \\
		axonal & K{\_}Pst & gK{\_}Pstbar & uniform & \SI{}{\siemens\per\cm\squared} & 0 & 1 \\
		axonal & K{\_}Tst & gK{\_}Tstbar & uniform & \SI{}{\siemens\per\cm\squared} & 0 & 0.1 \\
		axonal & SK{\_}E2 & gSK{\_}E2bar & uniform & \SI{}{\siemens\per\cm\squared} & 0 & 0.1 \\
		axonal & SKv3{\_}1 & gSKv3{\_}1bar & uniform & \SI{}{\siemens\per\cm\squared} & 0 & 2 \\
		axonal & Ca{\_}HVA & gCa{\_}HVAbar & uniform & \SI{}{\siemens\per\cm\squared} & 0 & 0.001 \\
		axonal & Ca{\_}LVAst & gCa{\_}LVAstbar & uniform & \SI{}{\siemens\per\cm\squared} & 0 & 0.01 \\
		axonal & CaDynamics{\_}E2 & gamma & uniform &  & 0.0005 & 0.05 \\
		axonal & CaDynamics{\_}E2 & decay & uniform &  \SI{}{\milli\second} & 20 & 1000 \\
		somatic & NaTs2{\_}t & gNaTs2{\_}tbar & uniform & \SI{}{\siemens\per\cm\squared} & 0 & 1 \\
		somatic & SKv3{\_}1 & gSKv3{\_}1bar & uniform & \SI{}{\siemens\per\cm\squared} & 0 & 1 \\
		somatic & SK{\_}E2 & gSK{\_}E2bar & uniform & \SI{}{\siemens\per\cm\squared} & 0 & 0.1 \\
		somatic & Ca{\_}HVA & gCa{\_}HVAbar & uniform & \SI{}{\siemens\per\cm\squared} & 0 & 0.001 \\
		somatic & Ca{\_}LVAst & gCa{\_}LVAstbar & uniform & \SI{}{\siemens\per\cm\squared} & 0 & 0.01 \\
		somatic & CaDynamics{\_}E2 & gamma & uniform &  & 0.0005 & 0.05 \\
		somatic & CaDynamics{\_}E2 & decay & uniform &  \SI{}{\milli\second} & 20 & 1000 \\
		\midrule
		Location & Mechanism & Parameter name & Distribution & Units & Value & \\
		\midrule
		global &  & v{\_}init &  &  \SI{}{\milli\volt} & -65 &   \\
		global &  & celsius &  &  \SI{}{\celsius} & 34 &   \\
		all &  & g{\_}pas & uniform & \SI{}{\siemens\per\cm\squared} & 3e-05 &   \\
		all &  & e{\_}pas & uniform &  \SI{}{\milli\volt} & -75 &   \\
		all &  & cm & uniform &  \SI{}{\micro\farad\per\cm\squared} & 1 &   \\
		all &  & Ra & uniform &  \SI{}{\ohm\cm} & 100 &   \\
		apical &  & ena & uniform &  \SI{}{\milli\volt} & 50 &   \\
		apical &  & ek & uniform &  \SI{}{\milli\volt} & -85 &   \\
		apical &  & cm & uniform &  \SI{}{\micro\farad\per\cm\squared} & 2 &   \\
		somatic &  & ena & uniform &  \SI{}{\milli\volt} & 50 &   \\
		somatic &  & ek & uniform &  \SI{}{\milli\volt} & -85 &   \\
		basal &  & cm & uniform &  \SI{}{\micro\farad\per\cm\squared} & 2 &   \\
		axonal &  & ena & uniform &  \SI{}{\milli\volt} & 50 &   \\
		axonal &  & ek & uniform &  \SI{}{\milli\volt} & -85 &   \\
		basal & Ih & gIhbar & uniform & \SI{}{\siemens\per\cm\squared} & 8e-05 &   \\
		apical & Ih & gIhbar & exp & \SI{}{\siemens\per\cm\squared} & 8e-05 &   \\
		somatic & Ih & gIhbar & uniform & \SI{}{\siemens\per\cm\squared} & 8e-05 &   \\
		\bottomrule

\end{tabular}
                                                              
\end{table}

An important aspect of this neuron model is the non-uniform distribution of certain ion channel conductances.
For example, the h-channel conductance is specified to increase exponentially with distance from the soma \citep{ihexp}, as follows.

\begin{lstlisting}
soma_loc = ephys.locations.NrnSeclistLocation(
	seclist_name='somatic',
	seclist_index=0,
	seg_x=0.5)

exponential_scaler = ephys.parameterscalers.NrnDistanceScaler(
	origin=soma_loc,
	distribution='(-0.8696 + 2.087*math.exp(({distance})*0.0031))*{value}')

parameter = ephys.parameters.NrnRangeParameter(
	name='gIhbar_Ih.apical',
	param_name='gIhbar_Ih'
	value_scaler=exponential_scaler,
	value=8e-5,
	frozen=True,
	locations=[apical_loc]))
\end{lstlisting}

\subsubsection{Protocols}

During the optimisation, the model is evaluated using three square current step stimuli applied and recorded at the soma.
For these protocols, a holding current is also applied during the entire stimulus, the amplitude of which is the same as was used in the \emph{in vitro} experiments to keep the cell at a standardised membrane voltage before the step current injection.

Another stimulus protocol checks for a backpropagating action potential (\textit{bAP}) by stimulating the soma with a very short pulse, and measuring the height and width of the bAP at a location of \SI{660}{\micro\metre} and \SI{800}{\micro\metre} from the soma in the apical dendrite.  It is specified as follows.

\begin{lstlisting}
for loc_name, loc_distance in [('dendloc1', 660), ('dendloc2', 800)]:
	loc = ephys.locations.NrnSomaDistanceCompLocation(
								name=loc_name,
								soma_distance=loc_distance)
	recording = nrpel.recordings.CompRecording(
					name='bAP.%s.v' % (loc_name),
					location=loc)
\end{lstlisting}

\subsubsection{Objectives}

For each of the four stimuli defined above, a set of eFeatures is calculated (Table \ref{featuretable}).
These are then compared with the same features extracted from experimental data.
As described in \cite{nmc}, experiments were performed that applied these and other protocols to L5PCs \emph{in vitro}.
For these cells, the same eFeatures were extracted, and the mean $\mu_{exp}$ and standard deviation $\sigma_{exp}$ calculated.
The bAP target values are extracted from \citet{bap}.

For every feature value $f_{model}$, one objective value is calculated:
$$objective = \left|\dfrac{\mu_{exp} - f_{model}}{\sigma_{exp}}\right|$$

\begin{table}
\textbf{\refstepcounter{table}\label{featuretable} Table \arabic{table}.}{ List of eFeatures for L5PC example. 
Locations \emph{dend1} and \emph{dend2} are respectively \SI{660}{\micro\metre} and \SI{800}{\micro\metre} from the soma in the apical trunk.
Depending on the eFeature type the units can be \SI{}{\milli\volt} or \SI{}{\milli\second}}.

\begin{tabular}{llllll}
	\toprule
		Stimulus & Location & eFeature & Mean & Std \\ 
		\midrule
		Step1 & soma & AHP{\_}depth{\_}abs & -60.3636 & 2.3018 \\
		 &  & AHP{\_}depth{\_}abs{\_}slow & -61.1513 & 2.3385 \\
		 &  & AHP{\_}slow{\_}time & 0.1599 & 0.0483 \\
		 &  & AP{\_}height & 25.0141 & 3.1463 \\
		 &  & AP{\_}width & 3.5312 & 0.8592 \\
		 &  & ISI{\_}CV & 0.109 & 0.1217 \\
		 &  & adaptation{\_}index2 & 0.0047 & 0.0514 \\
		 &  & doublet{\_}ISI & 62.75 & 9.6667 \\
		 &  & mean{\_}frequency & 6 & 1.2222 \\
		 &  & time{\_}to{\_}first{\_}spike & 27.25 & 5.7222 \\
		Step2 & soma & AHP{\_}depth{\_}abs & -59.9055 & 1.8329 \\
		 &  & AHP{\_}depth{\_}abs{\_}slow & -60.2471 & 1.8972 \\
		 &  & AHP{\_}slow{\_}time & 0.1676 & 0.0339 \\
		 &  & AP{\_}height & 27.1003 & 3.1463 \\
		 &  & AP{\_}width & 2.7917 & 0.7499 \\
		 &  & ISI{\_}CV & 0.0674 & 0.075 \\
		 &  & adaptation{\_}index2 & 0.005 & 0.0067 \\
		 &  & doublet{\_}ISI & 44.0 & 7.1327 \\
		 &  & mean{\_}frequency & 8.5 & 0.9796 \\
		 &  & time{\_}to{\_}first{\_}spike & 19.75 & 2.8776 \\
		Step3 & soma & AHP{\_}depth{\_}abs & -57.0905 & 2.3427 \\
		 &  & AHP{\_}depth{\_}abs{\_}slow & -61.1513 & 2.3385 \\
		 &  & AHP{\_}slow{\_}time & 0.1968 & 0.0112 \\
		 &  & AP{\_}height & 19.7207 & 3.7204 \\
		 &  & AP{\_}width & 3.5347 & 0.8788 \\
		 &  & ISI{\_}CV & 0.0737 & 0.0292 \\
		 &  & adaptation{\_}index2 & 0.0055 & 0.0015 \\
		 &  & doublet{\_}ISI & 22.75 & 4.14 \\
		 &  & mean{\_}frequency & 17.5 & 0.8 \\
		 &  & time{\_}to{\_}first{\_}spike & 10.5 & 1.36 \\
		bAP & dend1 & AP{\_}amplitude{\_}from{\_}voltagebase & 45 & 10 \\
		 & dend2 & AP{\_}amplitude{\_}from{\_}voltagebase & 36 & 9.33 \\
		 & soma & AP{\_}height & 25.0 & 5.0 \\
		 &  & AP{\_}width & 2.0 & 0.5 \\
		 &  & Spikecount & 1.0 & 0.01 \\
		\bottomrule

\end{tabular}

\end{table}

\subsubsection{Optimisation}

\begin{figure}
	\begin{minipage}[b]{.45\linewidth}
	\begin{center}
		\includegraphics[scale=.35,keepaspectratio]{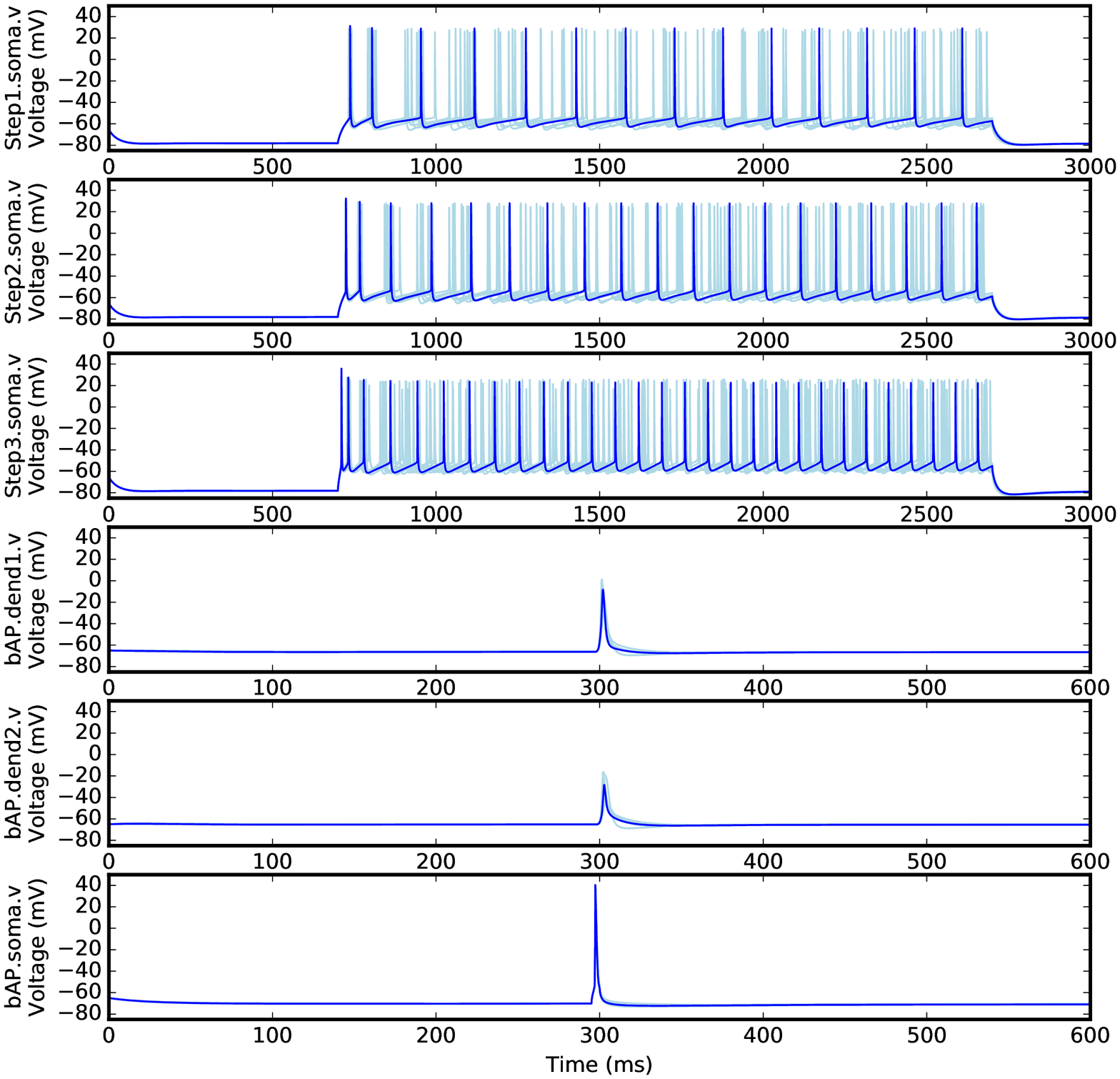}
	\end{center}
		\subcaption{Similar to Fig. \ref{fig:l5pc_release_model_responses}, with the top ten objective values found by BluePyOpt, and the best one plotted darker}\label{fig:l5pc_result}
	\end{minipage}	
 	\hspace{0.05\linewidth} 	
	\begin{minipage}[b]{.45\linewidth}
		\begin{center}
		\includegraphics[scale=.35,keepaspectratio]{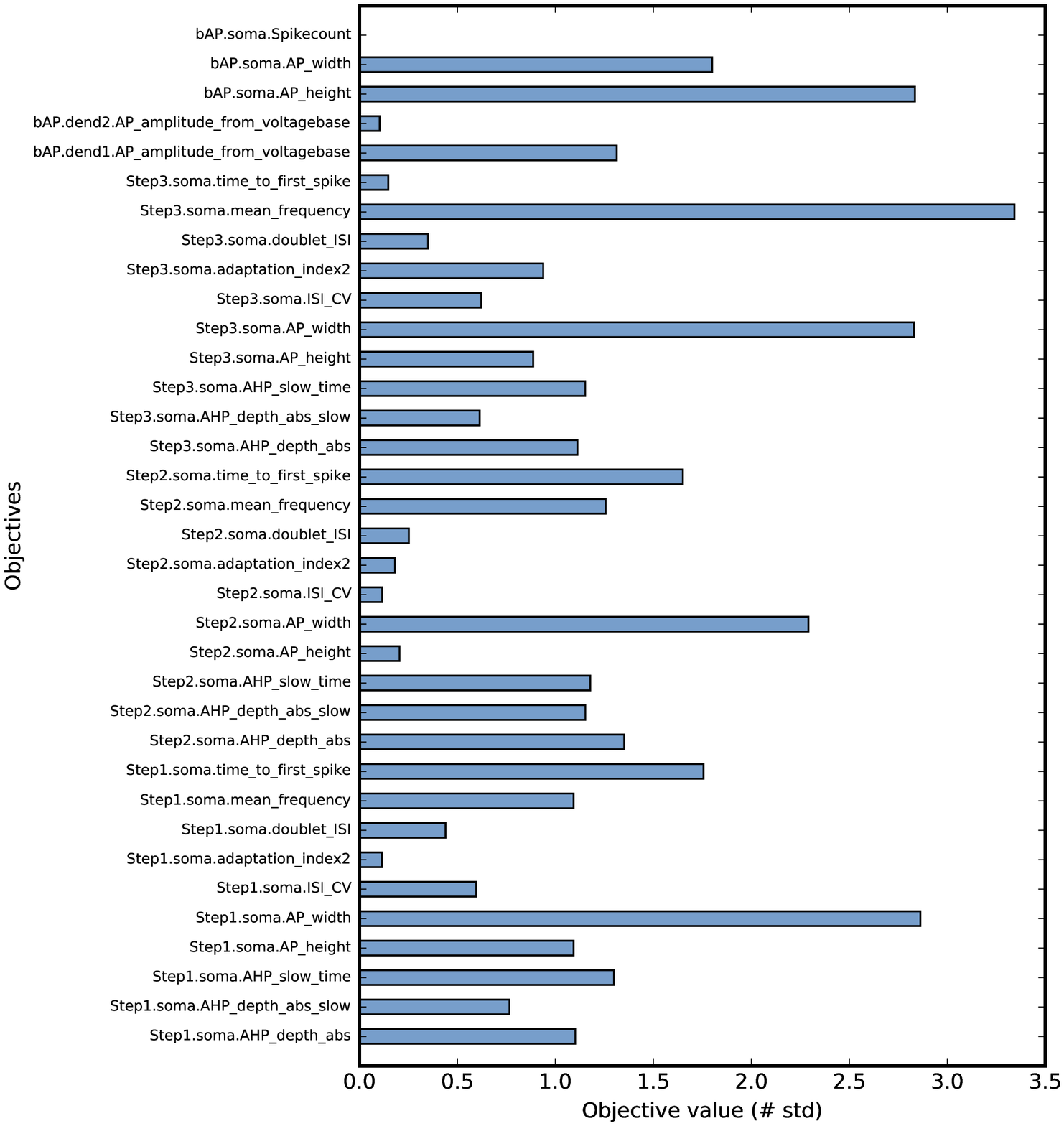}
		\end{center}
 		\subcaption{Objective scores for the best objective values found by BluePyOpt.}\label{fig:l5pc_model}
 	\end{minipage}
	
	\medskip	 	
 	
	\begin{minipage}[b]{.45\linewidth}
		\includegraphics[scale=.32,keepaspectratio]{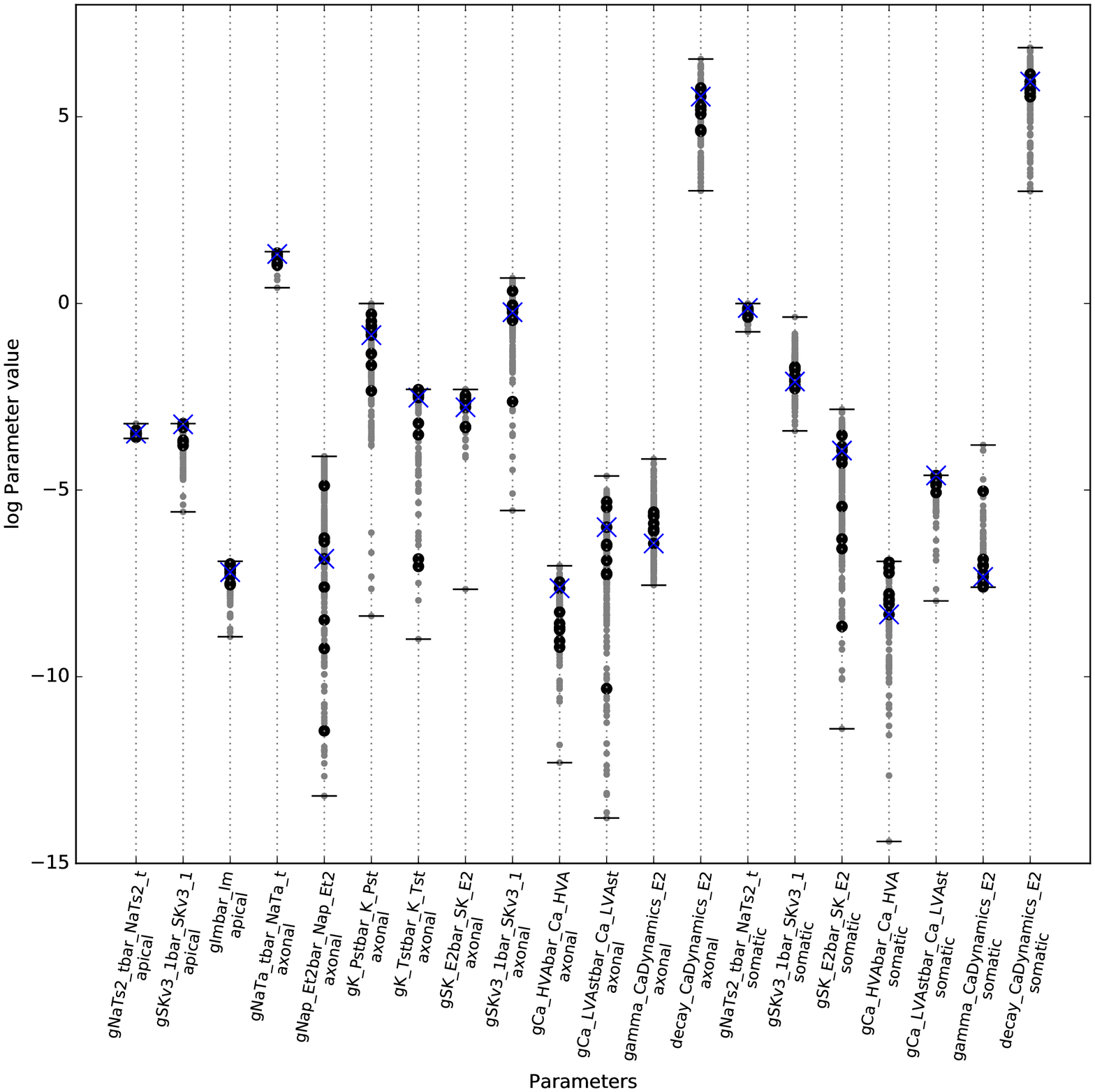}
		\subcaption{Parameter diversity in final solution. Parameter values for best (\emph{blue crosses}) and 10 best individuals (\emph{black dots}) and all individuals with all objectives below 5 (\emph{grey dots}).}\label{fig:l5pc_diversity}
	\end{minipage}
	 \hspace{0.05\linewidth} 	
	\begin{minipage}[b]{.45\linewidth}
		\includegraphics[scale=.35,keepaspectratio]{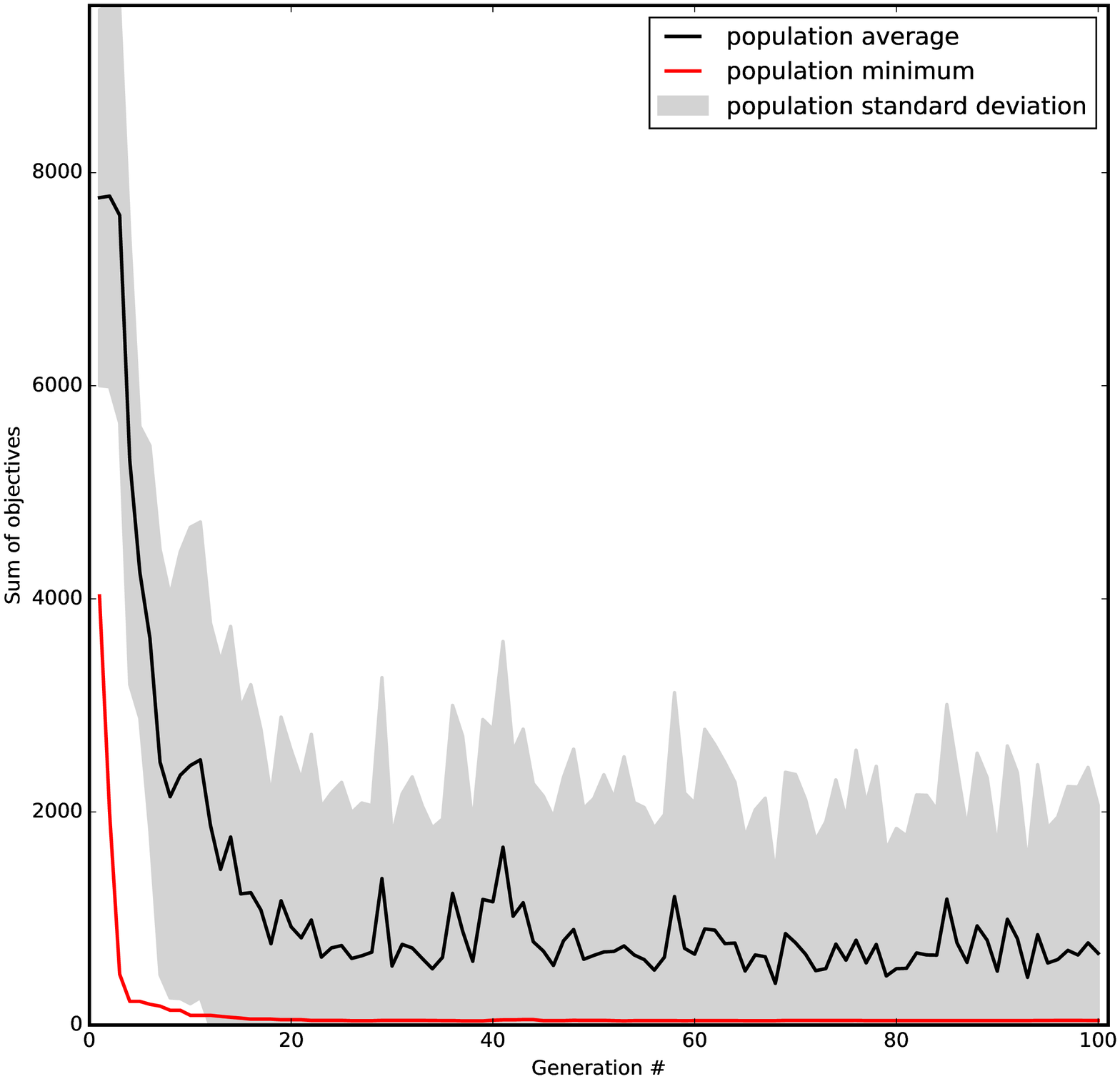}
		\subcaption{Evolution of the L5PC optimisation that found the model in panel \ref{fig:l5pc_result}. Plot shows the minimal, maximal and average scores found in the consecutive generations of the evolutionary algorithm.}\label{fig:l5pc_bpop_evolution}
	\end{minipage}
	\caption{Results of optimising L5PC model using BluePyOpt.}\label{fig:l5pc_bpop}
\end{figure}

\begin{figure}
    \begin{center}
		\includegraphics[scale=.5,keepaspectratio]{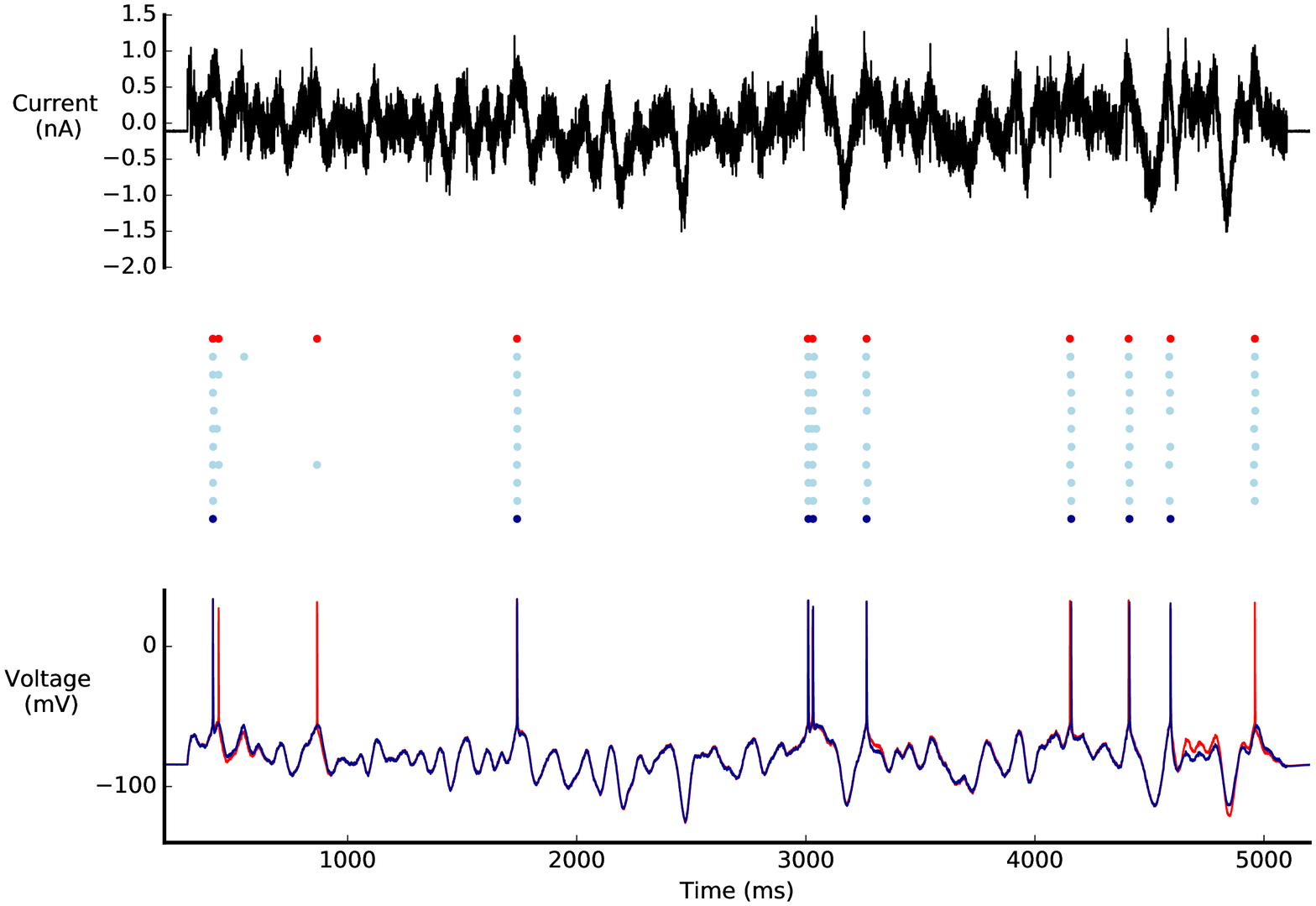}
	\end{center}
	\caption{Comparison of L5PC model solutions found by BluePyOpt to reference model. \emph{Top}: Gaussian noise current injected in the models. 
	\emph{Middle}: Raster plot of model responses to noise current injection.
	\emph{Bottom}: Voltage response of the models to noise current injection. In \emph{red}, model parameters from \cite{nmc} model, in \emph{light blue} the best 10 individuals found by BluePyOpt, in \emph{dark blue} the best individual. 
	Figure as in \cite{giffit}.}\label{fig:l5pc_valid}
\end{figure}

For the optimisation of this cell model we needed significantly more computing resources.
The goal was to find a solution that has objective values that are only several standard deviation away from the experimental mean.
For this we ran 200 generations with an offspring size of the genetic algorithm of 100 individuals.
The evaluation of these 100 individuals was parallelised over 50 CPU cores using SCOOP, and took a couple of hours to run. Figure \ref{fig:l5pc_bpop} shows the results of the optimisation, and Figure \ref{fig:l5pc_valid} shows a comparison of the optimised model to its reference under Gaussian noise current injection (not used during the optimisation).

\subsection{Spike-timing dependent plasticity (STDP) model}

The BluePyOpt framework was designed to be versatile and broadly applicable to a wide range of neuroscientific optimisation problems.
In this use case, we demonstrate this versatility by using BluePyOpt to optimise the parameters of a calcium-based STDP model \citep{graupner2012calcium} to summary statistics from in vitro experiments \citep{nevian2006spine}.
That is, we show how to fit the model to literature data, commonly reported just as mean and SEM of the amount of potentiation (depression) induced by one or more stimulation protocols.

In the set of experiments performed by \citet{nevian2006spine}, a presynaptic action potential (AP) is paired with a burst of three post-synaptic APs to induce either long-term potentiation (LTP) or long-term depression (LTD) of the postsynaptic neuron response. The time difference \(\Delta t\) between the presynaptic AP and the postsynaptic burst determines the direction of change: A burst shortly preceding the presynaptic AP causes LTD, with a peak at \(\Delta t = -50\) ms; conversely, a burst shortly after the presynaptic AP results in LTP, with a peak at \(\Delta t = +10\) ms \citep{nevian2006spine}. 

The model proposed by \citet{graupner2012calcium} assumes bistable synapses, with plasticity of their absolute efficacies governed by post-synaptic calcium dynamics.
That is, each synapse is either in an high-conductance state or a low-conductance state; potentiation and depression translate then into driving a certain fraction of synapses from the low-conductance state to the high-conductance state and vice versa; synapses switch from one state to another depending on the time spent by post-synaptic calcium transients above a potentiation (depression) threshold.
Following \citet{graupner2012calcium}, the model is described as

\begin{align}
\tau \frac{d\rho}{dt} &= - \rho(1-\rho)(\rho_\star-\rho) + \gamma_p(1-\rho)\Theta[c(t) - \theta_p]\\
                      & \qquad - \gamma_d \rho \Theta[c(t) - \theta_d] + \text{Noise}(t) \nonumber\\
\frac{dc}{dt} &= - \frac{c}{\tau_\text{Ca}} + C_\text{pre} \sum_i \delta(t-t_i-D) + C_\text{post} \sum_j \delta(t - t_j)
\end{align}
where \(\rho\) is the absolute synaptic efficacy, \(\rho_\star\) delimits the basins of attraction of the potentiated and depressed state, \(\gamma_p\) (\(\gamma_d\)) is the potentiation (depression) rate, \(\Theta\) is the Heaviside function, \(\theta_p\) (\(\theta_d\)) is the potentiation (depression) threshold, Noise\((t)\) is an activity dependent noise.
The postsynaptic calcium concentration is described by the process \(c\), with time constant \(\tau_\text{Ca}\).
\(C_\text{pre}\) is the calcium transient caused by a presynaptic spike with a delay \(D\) to account for the slow activation of NMDARs, while \(C_\text{post}\) is the calcium transient caused by a postsynaptic spike.

For periodic stimulation protocols, such as in \cite{nevian2006spine}, the synaptic transition probability can be easily calculated analytically \citep{graupner2012calcium}, allowing estimation of the amount of potentiation (depression) induced by the stimulation protocol without actually running any neuron simulations.
The amount of potentiation (depression) obtained with different protocols \emph{in vitro} become the objectives of the optimisation.

A small Python module \emph{stdputil} calculating this model is available in the example section on the BluePyOpt website.
To optimise this model, only an \emph{Evaluator} class has to be defined that implements an evaluation function:

\begin{lstlisting}
class GraupnerBrunelEvaluator(bpop.evaluators.Evaluator):
    def __init__(self):
        super(GraupnerBrunelEvaluator, self).__init__()
        # Graupner-Brunel model parameters and boundaries
        # From Graupner and Brunel (2012)
        self.graup_params = [('tau_ca', 1e-3, 100e-3),
                             ('C_pre', 0.1, 20.0),
                             ('C_post', 0.1, 50.0),
                             ('gamma_d', 5.0, 5000.0),
                             ('gamma_p', 5.0, 2500.0),
                             ('sigma', 0.35, 70.7),
                             ('tau', 2.5, 2500.0),
                             ('D', 0.0, 50e-3),
                             ('b', 1.0, 100.0)]

        self.params = [bpop.parameters.Parameter
                       (param_name, bounds=(min_bound, max_bound))
                       for param_name, min_bound, max_bound in self.
                       graup_params]

        self.param_names = [param.name for param in self.params]

        self.protocols, self.sg, self.stdev, self.stderr = \
            stdputil.load_neviansakmann()

        self.objectives = [bpop.objectives.Objective(protocol.prot_id)
                           for protocol in self.protocols]

    def get_param_dict(self, param_values):
        return gbParam(zip(self.param_names, param_values))
        
    def compute_synaptic_gain_with_lists(self, param_values):
        param_dict = self.get_param_dict(param_values)

        syn_gain = [stdputil.protocol_outcome(protocol, param_dict) \
                    for protocol in self.protocols]

        return syn_gain
                    
    def evaluate_with_lists(self, param_values):
        param_dict = self.get_param_dict(param_values)

        err = []
        for protocol, sg, stderr in zip(self.protocols, self.sg, self.stderr):
            res = stdputil.protocol_outcome(protocol, param_dict)

            err.append(numpy.abs(sg - res) / stderr)

        return err
\end{lstlisting}

\begin{figure}
	\begin{minipage}[b]{.45\linewidth}
		\includegraphics[scale=.3,keepaspectratio]{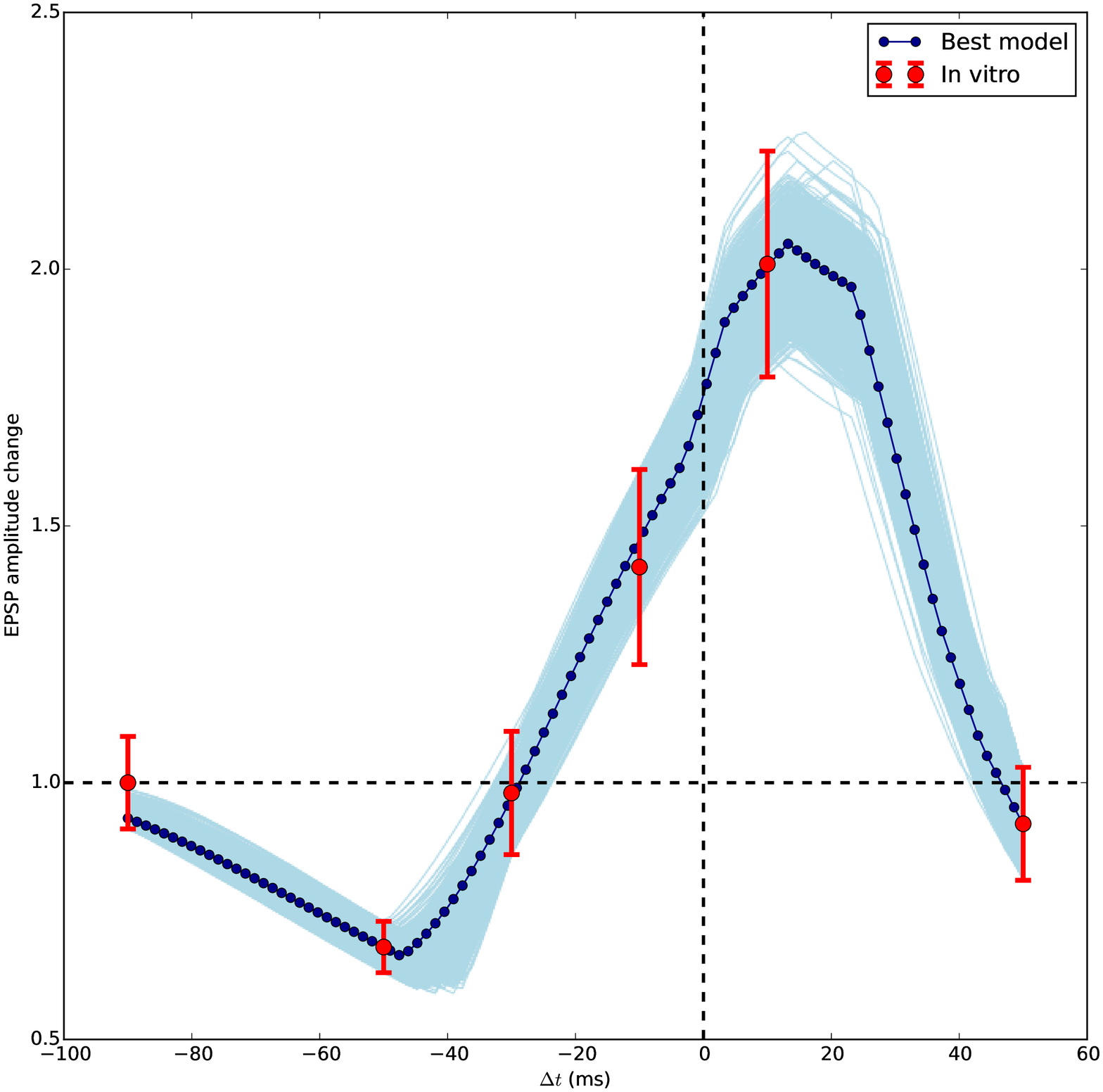}
		\subcaption{Comparison between model and experimental results; the models match the available \emph{in vitro} data and predict the outcome of the missing points. In \emph{light blue}, models generated by individuals having fitness values within one standard error of the mean from experimental \emph{in vitro} data. In \emph{dark blue}, best model, defined as the closest to all experimental data points. Experimental data from \citet{nevian2006spine} digitized using \citet{webdigi}.}\label{fig:stdp_model_curve}
	\end{minipage}
	\hspace{0.05\linewidth}
	\begin{minipage}[b]{.45\linewidth}
		\includegraphics[scale=.35,keepaspectratio]{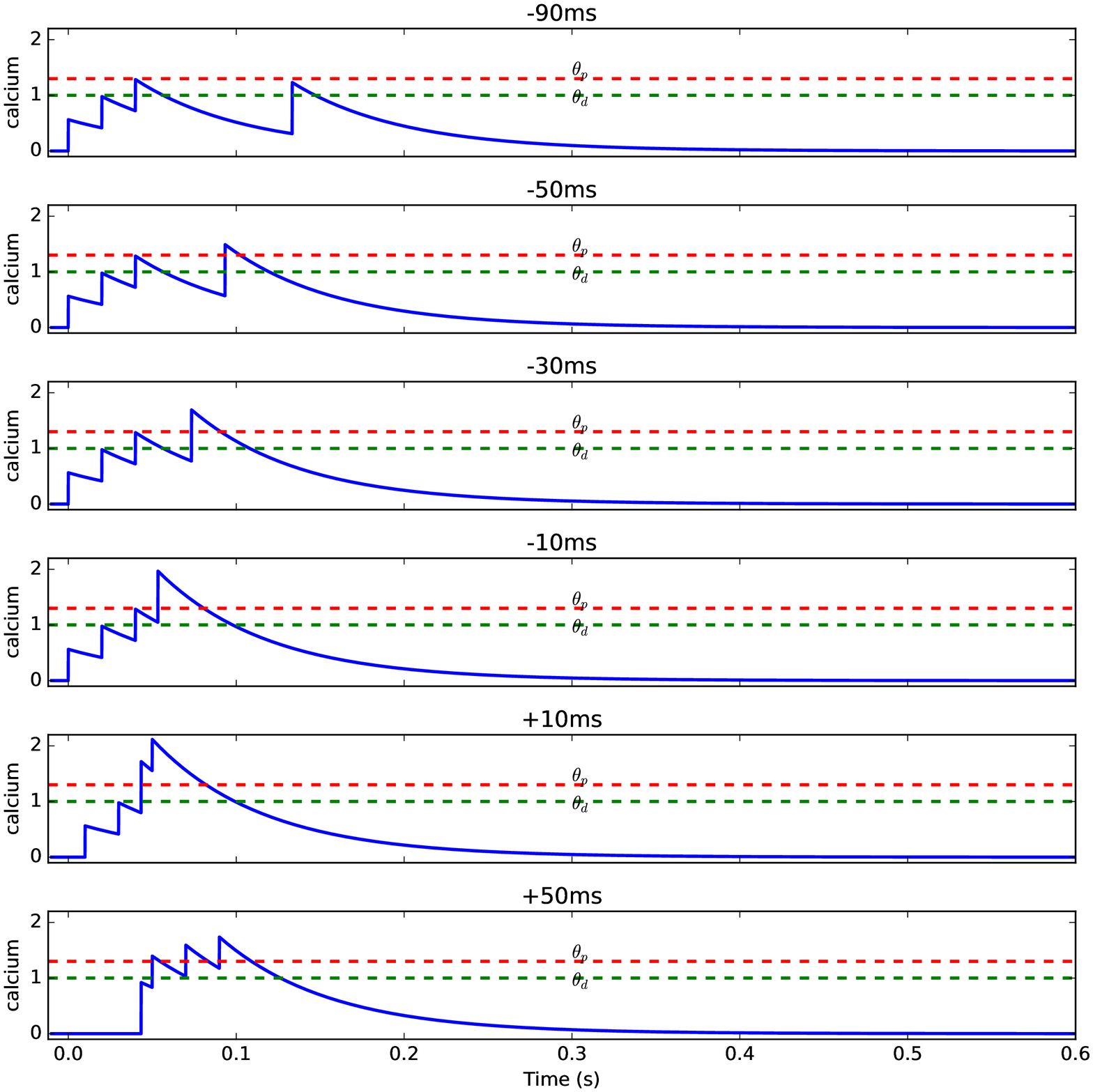}
		\subcaption{Calcium transients generated by the best model (panel a) for each stimulation protocol. Potentiation and depression thresholds, \(\theta_p\) and \(\theta_d\) respectively, are indicated by the dashed lines.}\label{fig:stdp_model_evolution}
	\end{minipage}
	
	\medskip	
	
	\begin{minipage}[b]{.45\linewidth}
		\includegraphics[scale=.3,keepaspectratio]{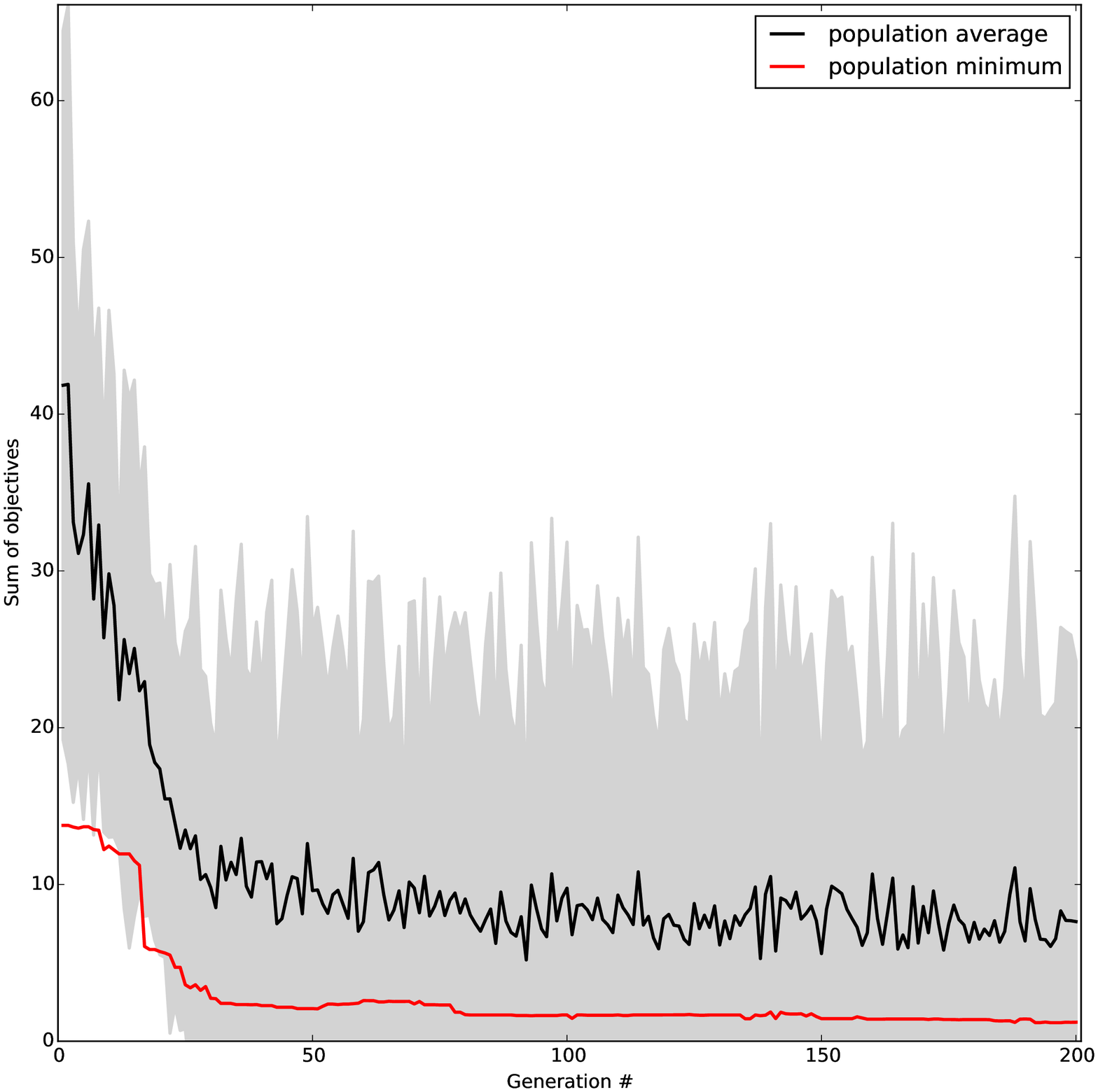}
		\subcaption{Evolution of the STDP optimisation that found the model in panel \ref{fig:stdp_model_curve}. 
		Minimal and average scores found in the consecutive generations of the evolutionary algorithm.}\label{fig:stdp_model_evoluti	on}
	\end{minipage}
	
	\caption{Results of STDP fitting.}\label{fig:stdp_model}
\end{figure}

With the evaluator defined, running the optimisation becomes as simple as:

\begin{lstlisting}
evaluator = GraupnerBrunelEvaluator()

opt = bpop.optimisations.DEAPOptimisation(GraupnerBrunelEvaluator())
results = opt.run(max_ngen=200)
\end{lstlisting}

Figure \ref{fig:stdp_model} shows the results of the optimisation.

\section{Discussion}

BluePyOpt was designed to be a state-of-the-art tool for neuroscientific model parameter search problems that is both easy to use for inexperienced users, and versatile and broadly applicable for power users.
Three example use cases were worked through in the text to demonstrate how BluePyOpt serves each of these user communities.

From a software point of view, this dual goal was achieved by an object oriented architecture which abstracts away the domain-specific complexities of search algorithms and simulators, while allowing extension and modification of the implementation and settings of an optimisation.
Python was an ideal implementation language for such an architecture, with its very open and minimal approach to extending existing implementations.
Object oriented programming allows users to define new subclasses of existing BluePyOpt API classes with different implementations.
The \textit{duck typing} of Python allows parameters and objectives to have any kind of type, e.g. they don't have to be floating point numbers.
In extreme cases, function implementations can even be overwritten at run time by \textit{monkey patching}.
These features of Python gives extreme flexibility to the user, which will make BluePyOpt applicable to many use cases.

A common issue arising for users of optimisation software is the configuration of computing infrastructure.
The fact that BluePyOpt is coded in Python, an interpreted language, and provides Ansible scripts for its installation, makes straightforward to run on diverse computing platforms.
This will give the user the flexibility to pick the computing infrastructure which best fits their needs, be it their desktop computer, university cluster or temporarily rented cloud infrastructure, such as offered by Amazon Web Services.

This present paper focuses on the use of BluePyOpt as an optimisation tool.
It is worth noting that the application domain of BluePyOpt needn't remain limited to this.
The ephys model abstraction can also be used in validation, assessing generalisation, and parameter sensitivity analyses.
E.g. when applying a map function to an ephys model evaluation function which takes as input a set of morphologies, one can measure how well the model generalises when applied to different morphologies.
The present paper expressly does not touch on issues of generalization power, overfitting, or uniqueness of solution.
It is worth now making a few points on the latter.  While BluePyOpt could successfully optimise the three examples, Figures \ref{fig:simplecell_trip} and \ref{fig:l5pc_diversity} show a diversity of solutions giving good fitness values.
That is, for these neuron model optimisation problems, the solutions found are non-unique.
This is compatible with the observation that Nature itself also utilises various and non-unique solutions to provide the required “phenotype” \citep{Schulz, marder}.
For other problems solutions could be unique, making BluePyOpt useful e.g. for extracting parameters for models of synapse dynamics \citep{Fuhrmann140}.
  
Of course BluePyOpt is not the only tool available to perform parameter optimisations in neuroscience \citep{druckmann, neurofitter, emoo, optimizer, carlson, giffit}.
Some tools provide a Graphical User Interface (GUI), other tools are written in other languages, or use different types of evaluation functions or search algorithms.
We explicitly didn't make a detailed comparison between BluePyOpt and other tools because many of these tools are developed for specific and non-overlapping applications, making a systematic comparison difficult. This suggests perhaps BluePyOpt's greatest strength, its broad applicability relative to previous approaches.

While BluePyOpt significantly reduces the domain specific knowledge required to employ parameter optimisation strategies, some thought from the user in setting up their problem is still required.  For example, BluePyOpt does in principle allow brute force optimisation of all parameters of the L5PC model example, including channel kinetics parameters and passive properties, but such an approach would almost certainly be unsuccessful.  Moreover, when it comes to assessing fitness of models, care and experience is also required to avoid the optimisation getting caught in local minima, or cannibalizing one objective for another.  For neuron models for example, feature-based approaches coupled with multi-objective optimisation strategies have proven especially effective \citep{druckmann}.  Indeed, even the stimuli and features themselves can be optimised on theoretical grounds to improve parameter optimisation outcomes \citep{druckmann_plos}.  For these reasons, an important companion of BluePyOpt will be a growing library of working optimisation examples developed by domain experts for a variety of common use cases, to help inexperienced users quickly adopt a working strategy most closely related to their specific needs.

As these examples library grows, so too will the capabilities of BluePyOpt evolve.  Some improvements planned for the future include the following:
\begin{itemize}
\item[] \textbf{Support for multi-stage optimisations} allowing for example the passive properties of a neuron to optimised in a first stage, prior to optimising the full-active dendritic parameters in a second phase
\item[] \textbf{Embedded optimisation} allowing for example an optimisation of a ``current at rheobase'' feature requiring threshold detection during the optimisation using e.g. a binary search.  Also, for integrate-and-fire models such as the adapting exponential integrate-and-fire \citep{Brette3637}, a hybrid of a global stochastic search and local gradient descent has been shown to be a competitive approach \citep{Jolivet2008}
\item[] \textbf{Fast pre-evaluation of models} to exclude clearly bad parameters before computation time is wasted on them
\item[] \textbf{Support for evaluation time-outs} to protect against optimisations getting stuck in long evaluations, for example when using NEURON's CVODE solver, which can occasionally get stuck at excessively high resolutions.
\item[] \textbf{Support for explicit units} to make optimisation scripts more readable, and sharing with others less error prone.
\end{itemize}

Although parameter optimisations can require appreciable computing resources, the ability to share the code of an optimisation through a light-weight script or ipython notebook using BluePyOpt will improve reproducibility in the field.
It allows for neuroscientists to exchange code and knowledge about search algorithms that perform well for particular models.
In the future, making it possible for users to read and write model descriptions from community standards \citep{neuroml, nineml}, could further ease the process of plugging in a model into a BluePyOpt optimisation.
By providing the neuroscientific community with BluePyOpt, an open source tool to optimise model parameters in Python which is powerful, easy to use and broadly applicable, we hope to catalyse community uptake of state-of-the-art model optimisation approaches, and encourage code sharing and collaboration.

\section*{Downloads}

The source code of BluePyOpt, the example scripts and cloud installation scripts are available on Github at \url{https://github.com/BlueBrain/BluePyOpt}, the former under the GNU Lesser General Public License version 3 (LGPLv3), and the latter two under a BSD license.

\section*{Disclosure/Conflict-of-Interest Statement}

The authors declare that the research was conducted in the absence of any commercial or financial relationships that could be construed as a potential conflict of interest.

\section*{Author Contributions}

WVG, MG and JDC designed the software and contributed code.
WVG, GC, MG and CR designed the examples and contributed code.
WVG, EM, MG and GC wrote the manuscript.
All: Conception and design, drafting and revising, and final approval.

\section*{Acknowledgments}

We wish to thank Michael Graupner for his support with the implementation of the calcium-based STDP model \citep{graupner2012calcium} and for fruitful discussions, and Elisabetta Iavarone for testing the cloud installation functionality.

\textit{Funding\text{\rm :}} The work was supported by funding from the EPFL to the Laboratory of Neural Microcircuitry (LNMC) and funding from the ETH Domain for the Blue Brain Project (BBP).
Additional support was provided by funding for the Human Brain Project from the European Union Seventh Framework Program (FP7/2007- 2013) under grant agreement no. 604102 (HBP).
The BlueBrain IV BlueGene/Q and Linux cluster used as a development system for this work is financed by ETH Board Funding to the Blue Brain Project as a National Research Infrastructure and hosted at the Swiss National Supercomputing Center (CSCS).

\bibliographystyle{chicago}
\bibliography{optpaper-arxiv}

\newpage

\end{document}